\documentclass[preprint]{aastex631}

\usepackage{color}
\newcommand{\ME}{M_{\rm E}}

\newcommand{\MJ}{M_{\rm J}}
\newcommand{\RJ}{R_{\rm J}}

\newcommand{\Tone}{T_{\rm 1\,\rm bar}}
\newcommand{\PHe}{P_{\rm He}}
\newcommand{\PZ}{P_{\rm Z}}
\newcommand{\Zatm}{Z_{\rm atm}}
\newcommand{\Zdeep}{Z_{\rm deep}}

\shorttitle{Jupiter models with OSL}
\shortauthors{Nettelmann \& Fortney}

\graphicspath{{./}{./}}

\begin{document}
\fontsize{11}{15}\selectfont

\title{Jupiter's Interior with an Inverted Helium Gradient}

\author[0000-0002-1608-7185]{N. Nettelmann}
\affiliation{Department of Astronomy \& Astrophysics, University of California, Santa Cruz, CA 95064, USA}

\author[0000-0002-9843-4354]{J.~J. Fortney}
\affiliation{Department of Astronomy \& Astrophysics, University of California, Santa Cruz, CA 95064, USA}

\correspondingauthor{N. Nettelmann}
\email{nadine.nettelmann@gmx.de}

\begin{abstract}

Jupiter's gravity field observed by NASA's Juno spacecraft indicates that the density in the 10--100 GPa 
region is lower than one would expect from a H/He adiabat with 0.5-5$\times$ solar water abundance as has been
observationally inferred in Jupiter's atmosphere, supported by the 2--4$\times$ solar enrichment in the heavy 
noble gases and other volatiles observed by the Galileo entry probe.

Here, we assume that Jupiter's envelope harbors a radiative window at  $\sim$0.975--0.99$\RJ$. 
This outer stable layer (OSL) delays particle exchange and accelerates the cooling of the deep interior. 
Consequently, the He-depletion at the Mbar-level where H/He phase separation occurs would be 
stronger than seen in the atmosphere.

We find that the inverted He-gradient across the OSL leads to atmospheric heavy element abundances 
that are up to $\Delta \Zatm =0.03$ $(+2\times$ solar) higher than for adiabatic models. 
With an additional inverted Z-gradient, $\Zatm$ up to $3\times$solar is possible. 
Models with 1$\times$ solar $\Zatm$ have a dilute core confined to the inner 0.2--0.3$\MJ$
(0.4--0.5$\RJ$), smaller than in adiabatic models. 
Models with 3$\times$ solar $\Zatm$ have a largely homogeneous-Z interior at 1$\times$ solar.

The low observed atmospheric Ne/He ratio suggests that Ne is transported through the OSL 
as efficiently as He is and at an enhanced diffusivity as is characteristic of double diffusive convection.
Better knowledge of the H/He-EOS in the 10---100 GPa region and of the H/He phase diagram is 
needed to understand Jupiter's interior structure.

\end{abstract}

\section{Introduction}
 
The planet Jupiter is a giant sample of protosolar nebula material. It harbors information 
on the origin of the solar system and how the planets formed \citep{Bolton17,Helled22}. 
Over its 4.5 billion of years in orbit around Sun, Jupiter's atmospheric composition may have 
been influenced by pollution from small bodies that migrate through the solar system and eventually
hit a planet \citep{Howard23invZ,Muller24}. 
While overall conserved, the distribution of the elements in Jupiter's interior may have changed as 
cooling initiates phase transitions. For Jupiter, the most relevant possible phase transition 
is the separation of the protosolar gas into a He-poor phase and a He-rich phase in a dense, 
metallic-hydrogen environment \citep{SS77a, SS77b}. To reveal information about its bulk composition, 
composition distribution, and the behavior of dense H/He requires interior structure models 
that are consistent with the relevant observational constraints and are built on experimentally 
confirmed equations of state (EOS) of the major constituents hydrogen (H) and helium (He). 

With the availability of first-principles based EOSs, such as the DFT-MD method (\citealp[e.g.,][]{Collins95}), 
and their experimental verification at benchmark points in shock \citep{KD17} or quasi-isentropic 
(\citealp[e.g.,][]{Fortov07}) compression experiments, and with the accurate gravitational harmonics 
measurements by the Juno spacecraft \citep{Durante20}, Jupiter models are now constrained better than 
ever and thus one may think that Jupiter's interior and formation were now well understood.
  
However, current constraints are so tight that is has become a challenge to find consistent 
Jupiter models at all. In particular, the gravitational harmonic $J_4$ implies a lower density 
in the 10---100 GPa region \citep{N21,Howard23} than what was allowed within the previous uncertainty range. 
The low density may indicate a sub-solar atmospheric metallicity, but for some H/He-EOSs, 
even a pure H/He adiabat seems to be too dense.
The low density implied by the gravity data places tight constraints on the H/He-EOS. 
Moreover, this challenge questions the classical assumption of an adiabatic interior \citep{Hubbard68}, an assumption 
that otherwise has been highly successful in explaining Jupiter's luminosity \citep{Fortney11,N12,MF20}. 

A little relief has recently come from Juno measurements of Jupiter's thermal emission, which suggest 
that the adiabat underneath the weather layer may be warmer \citep{Li24} and thus less dense
than an adiabat constrained by the Galileo entry probe measurement of a low 1-bar temperature of $\Tone=$166.1 K.
The density reduction along a H/He adiabat due to a few ($\sim$8) K warmer troposphere is 
readily balanced by a small (0.5$\times$ solar) increment in atmospheric metallicity. 
For most H/He-EOS, this is not enough to let the models pass a 1$\times$ solar atmospheric metallicity threshold,
noting that Jupiter's true atmospheric metallicity could even be 3$\times$ solar or higher. 
This creates a significant tension. 
The magnitude of this tension will become more evident when Juno's major scientific goal 
of a global mean water abundance determination \citep{Bolton17} is completed.

The Jupiter models that currently best fit the data are based on mainly three different H/He-EOSs: 
the MH13 H/He-EOS, which includes non-ideal mixing contributions to the entropy for a single 
He-concentration and otherwise assumes linear mixing between He and H/He \citep{MH13,Militzer24}; 
second, the HG23+MLS22 EOS which includes a He-concentration-dependent non-ideal entropy of mixing 
term that is applied to the linear mixing of the CMS19-He EOS \citep{CMS19} and the MLS22 H-EOS \citep{Howard23}; 
and third, REOS.3, which assumes linear mixing between H, He, and heavy elements and is based on 
H and He-EOSs that differ in the sub-Mbar region from the SCvH-EOS \citep{Becker14,N17}. 
REOS-adiabats, however, can be subject to uncertainty due to thermodynamic integration over a 
not fully thermodynamically consistent EOS table.

The Jupiter models that best fit the observational data on gravity and 1-bar temperature are characterized 
by a He-poor/He-rich transition pressure $\PHe$ in the $\sim$1 to 6-Mbar range \citep{N17,HG23,Militzer24}, 
and an $\sim 1\times$ solar atmospheric metallicity \citep{Militzer24,HG23}.

Atmospheric metallicities higher than 1.3$\times$ solar are out of reach by current interior models and, 
depending on the H/He-EOS used, even 1x solar is made possible only under rather bold assumptions.
These include a 5-15 K higher 1-bar temperature than observed \citep{Miguel22}, He-depletion that
extends deep into the dilute core \citep{Debras19}, a superadiabatic 10--100 GPa region \citep{Debras19},
or adiabat perturbations that act to reduce the density in the 1--100 GPa region by up to 10\% \citep{Howard23}.
\citet{N21} presented this tension problem unveiled by using the H/He-EoS as is, in that case the CMS19-EOS. 
They used the metallicity as a free parameter, with positive values increasing the density along the H/He-adiabat 
while negative values decreasing it. Assuming a nominal $\Tone$ of 166.1 K and $\PHe=2$ Mbar consistent
with the pressures where H/He-phase separation is predicted to occur by
all phase diagrams, they found solutions to the gravity field only for negative values of $\Zatm$.  
This means that the pure CMS19-EOS based H/He adiabat seems to be too dense.
Their answer to the question how much higher $\Tone$ would have to be to obtain 1x solar $\Zatm$ was 180 K,
in agreement with later extensive Bayesian statistical analysis results for the same H/He-EOS  
\citep{Miguel22,HG23}. 

In this work, we assume that Jupiter is not fully adiabatic  but that is harbors a radiative zone in the deep
atmosphere. We aim to quantify how much this assumption helps to increase the atmospheric-Z of the interior models. 
We find that a deviation from adiabaticity is most effective if such a region is placed far out in the planet
at a $\sim 1$ kbar-level. We furthermore find that this outer stable layer (OSL) is in a regime of 
double diffusive convection.
The sub-adiabaticity is assumed to balance an inverted He-gradient, and optionally an additional inverted Z-gradient. 
The simple idea behind these assumptions is that the abundances of He and heavy elements can be higher in the 
atmosphere where they are observed than down in the region where $J_4$ is most sensitive. Helium offers a huge 
reservoir. Removing He allows to add more heavy elements in this region and all the way up into the atmosphere.

Recently, there has been independent observational indication for a radiative zone in Jupiter's deep atmosphere.
One indication is related to the CO abundance at a few bars \citep{Belzard02}, which is high or low depending on what one compares it to.
Jupiter's tropospheric CO abundance is high compared to the expectation from equilibrium chemistry between C-O-H-bearing
molecules such as CO, H$_2$, H$_2$O, and CH$_4$, suggesting that CO from greater depths is transported upward and that the 
vertical mixing timescale is shorter than the chemical reaction timescale. If disequilibrium chemistry is accounted 
for \citep{Cavalie23}, Jupiter's tropospheric CO abundance actually appears low compared to the expectation for 
an 1-5$\times$ solar deep O/H abundance. \citet{Cavalie23} suggest that either the deep O/H is low, $\lesssim 0.3$ solar, 
or that the vertical mixing is reduced because of the presence of a radiative zone at pressures deeper 
than 0.6 kbar, or temperatures higher than 1000 K.
Another indication is related to the analysis of brightness temperature and limb darkening observed by Juno/MWR.
To explain those data, \citet{Bhatta23} conclude that the opacity at 100-200 bars must be enhanced compared to that
expected from ammonia and water, and that this enhancement results from free electrons. These are released by ionized
alkali metals deeper down. The alkali metal abundance needed to explain the free electron density would correspond to
only $10^{-5}$ to $10^{-2}$ times solar.
In the entire absence of alkali metals, a pronounced radiative window opens in the 1300-2700 K range 
\citep{Guillot94}, which, according to our Jupiter models, would be at 0.975-0.99 $\RJ$ just where we favor the location 
of an OSL.

If an OSL exists, it can in addition have the effect that in-falling material from inter-planetary
space does not get immediately mixed with the layer under the stable region: an additional inverted  
Z-gradient builds up. \citet{Muller24} model the evolution of Jupiter with atmospheric pollution. They find 
that an inverted Z-gradient between a 3$\times$ solar atmosphere and a 1$\times$ solar deeper interior can be 
stable against mixing over billions of years under their assumption of an outer radiative zone that 
develops if the opacity is reduced by a factor of 10 compared to the Rosseland mean. In present
Jupiter, this radiative zone would be located in the 1--10 kbar region. 
Our Jupiter model set-up is consistent with these estimates.

In Section 2 we display the $\Tone$--atmospheric-$Z$ phase space from observations and models. 
In Section 3 we explain the layered structure model. 
Results are presented in Section 4 including $\Zatm$, $Z$-profiles, He-depletion from H/He phase
separation and transport of He and Ne through the OSL.
The idea of an outer stable layer in Jupiter is not itself novel. In the Section \ref{sec:discuss} 
we discuss alternative proposals for an OSL. Section 6 summarizes the paper.

For this purpose we use the CD21-EOS \citep{CD21} and constrain the models by the observed values 
of $\Tone$, $J_2$, and $J_4$.

\section{The O/H--T--Z phase space} \label{sec:OTZ}

\subsection{T-1~bar measurements}

We are interested in the temperature at the 1-bar level, which for interior models commonly serves 
as the outer boundary condition for the $P$--$T$ profile that is assumed to extend adiabatically inward.
This is of course an idealized assumption as it ignores possible deviations from adiabaticity in the 
weather layer in the vertical direction and latitudinal variations to the temperature due to gravity darkening.
In the latter case, we find that the temperature difference between pole and equator can amount 1--3 K,
depending on the gravity darkening exponent used, here: 0.08--0.25 \citep{EspinozaLara11}.

The most accurate $\Tone$-measurement has been achieved by the Galileo entry probe, which measured
166.1$\pm 0.8\:$K in a near-equatorial hot spot \citep{Seiff1998}. However, hot spots may be particularly 
dry regions of the atmosphere, and the dry adiabatic P--T profile therein not be representative of the 
global profile if the atmosphere is wetter on average. The entry probe measurement has been applied as
the outer boundary constraint to many Jupiter structure models \citep{HM16,N21,Militzer24}, and it is consistent with the temperature of $165\pm 5\:$K derived from the Voyager occultation experiment \citep{Lindal81}.
However, inference of temperature from the occultation data relies on assumptions about the composition of the atmosphere at the altitudes where the electromagnetic signal emitted by the spacecraft passes through and becomes refracted. In the $\Tone$-determination from the Voyager 1 and 2 data by \citet{Lindal81}, the mean molecular weight $\mu$ was assumed to be constant with altitude and to have the He volume mixing ratio of $q_{\rm He}=0.11\pm 0.03$ ($Y_{\rm HHe}=0.198(50)$ in a Z-free H/He atmosphere) as observed by remote sensing with the Voyager/IRIS instrument \citep{Hanel79}. 
\citet{Gupta22} conducted a re-analysis of the Voyager occultation data by using the Galileo values
for He ($q_{\rm He}=0.1356$) and heavy element abundances, and updated refractivity data
to conclude a $+4\:$K warmer atmosphere at the 1-bar level. Notably, they find an up to 7 K difference 
between the latitudes of ingress and egress, and within the $1\sigma$ uncertainty a maximum 
$\Tone$ of 174 K at $12^{\circ}$S. 
While Juno radio occultation data acquired during the Extended Mission are currently being analyzed for the thermal structure at different latitudes and down to pressures of up to 0.5 bar \citep{Smirnova24}, \citet{Li24} analyzed the brightness temperatures and limb darkening measured
by Juno/MWR. They and found a preference for a super-adiabatic temperature profile across the water cloud layer as well as a deep $P$--$T$ profile that if extrapolated outward to the 1-bar level, would yield a virtual temperature of $169\pm 1.6$ K. A virtual temperature that is higher (lower) than the observed physical temperature, here the Galileo value of 166.1 K, means that the deep adiabat is warmer (colder), but the super- (sub-)adiabatic temperature gradient across the stable layer, here the water cloud layer, lets the 1-bar level appear different from an adiabatically outward extended deep adiabat. The described $\Tone$ estimates from observations are displayed in Figure \ref{fig:OTZ}a.

\subsection{O/H abundance measurements}

To determine the deep water abundance underneath the water cloud layer is a key science goal of the Juno mission. As water is the main carrier of oxygen in the deep atmosphere, the water abundance is often reported as an O/H ratio where H refers to all hydrogen particles, bound or not. 
Several methods have provided O/H estimates using different instruments aboard the Juno spacecraft. 

The Microwave Radiometer (MWR) measures the emission from Jupiter in six different channels which cover
wavelengths from 1.3 to 50 cm, or 22 to 0.6 GHz. The pressure where the emission is estimated to come 
from increases with the observed wavelength. The deepest levels where emission can be detected is 
100-250 bars in channel 1 \citep{Bolton17,Janssen17}. This is well below the bottom of the
water clouds, which for a nominal water abundance of 3$\times$ solar would be at 4--5 bars \citep{Atreya20}. 
The brightness temperature $T_b$ of Jupiter at these wavelengths is dominated by the absorption of 
ammonia and to a lesser extent, but increasing with pressure, also of water. Measurement of $T_b$ and its emission angle dependence in terms of a limb-darkening parameter $R(\theta)$ in the equatorial zone down to $\sim 30$ bars revealed a rather variable ammonia abundance. Nevertheless, assuming a moist adiabatic temperature profile allowed \citet{Li20} to constrain the deep ammonia abundance to 2.6--3.0$\times$ solar while the water abundance to 1--5$\times$ solar (1$\sigma$ uncertainties). However, with 0.1--7.5$\times$ solar, the $2\sigma$ uncertainty in the equatorial water abundance remains quite large. \citet{Li24} find that a super-adiabatic temperature gradient best explains the Juno MWR data for brightness temperature and limb darkening and that it would require a mean molecular weight gradient that would correspond to a deep O/H of 4.9$\times$ solar for a virtual 1-bar temperature of 169 K with an uncertainty range of 1.5--8$\times$ solar that arises from 
the uncertainty in the temperature gradient.  The O/H--virtual temperature relation is obtained from 3D simulations of moist convection and indicates a linear trend that, by extrapolation beyond 3$\times$ solar O/H yields the quoted high O/H abundance estimate, which is the highest of all estimates. However, for a true higher physical 1-bar temperature of $\sim 169\:$K as suggested by the re-analyzed Voyager data, the super-adiabaticity would weaken and permit lower a water abundance of 0.1--1x solar \citep{Li24}.

Weather phenomena in Jupiter's shallow atmosphere such as clouds, storms, and lightning provide alternative constraints on the water abundance. \citet{Bjoraker22} used the Keck telescope to observe Jupiter at $\sim 5\mu$m and found evidence of water clouds at a few bars. Although their inferred water abundance refers to the cloud top and is not directly representative of the deep water abundance, \citet{Cavalie24} suggest a deep water abundance of 0.5--2.5$\times$ solar from these measurements. 

Storms play an important role in transporting heat from the deep interior outward. 
Intensive storms that overshoot the radiative atmosphere require strong moist convection at their bottom
as a power source \citep{Cavalie24}. The intensity and duration of the storms is determined by the available 
amount of condensible species, especially of water. Numerical simulations for recent intense storms on 
Jupiter observed with HST and JunoCam predict a minimum water abundance of 1$\times$ solar \citep{Inurrigarro22}.

Lightning events are also associated with moist convection and water clouds. Fits to the frequency of 
lightning events at shallow depth of 1-2 bars observed by Galileo and Juno are consistent with a
subsolar O/H \citep{Aglyamov21}. Lightning events at greater depth may point to higher water abundances 
there, although the exact lightning depth--water abundance relation also depends on processes like 
the deposition of water by hail at greater depth than the condensation level, and the evaporation of 
ice crystals by collisions \citep{Aglyamov23}.    
Overall, \citet{Cavalie24} conclude a minimum O/H of 0.5$\times$ solar from these weather phenomena.

Underneath the water clouds, methane and water are believed to be in thermochemical
equilibrium with carbonmonoxide (CO) and hydrogen. This is not the case in the high atmosphere. 
With an observed abundance there of 0.9 ppm, CO is by orders of magnitude more abundant than suggested by 
equilibrium chemistry. This information has been used to infer the deep water 
abundance by means of 1D-thermochemical and diffusion modelling and a reaction network that accounts
for disequilibrium  chemistry. Assuming furthermore that the region between where CO and H$_2$O are in 
equilibrium and that probed by Galileo is convective and can be described by a 
vertically constant high Eddy diffusion coefficient, \citet{Cavalie23} infer a low deep water 
abundance of 0.3$\times$ solar. Assuming alternatively the presence of a radiative region at 1400-2200 K 
and representing this region of limited mixing by a reduced Eddy diffusion coefficient, \citet{Cavalie23} 
obtain 1$\times$ solar O/H, although they emphasize such a solution is non-unique.  

\begin{figure*}
\begin{minipage}{0.96\textwidth}
\rotatebox{270}{\includegraphics[width=0.40\textwidth]{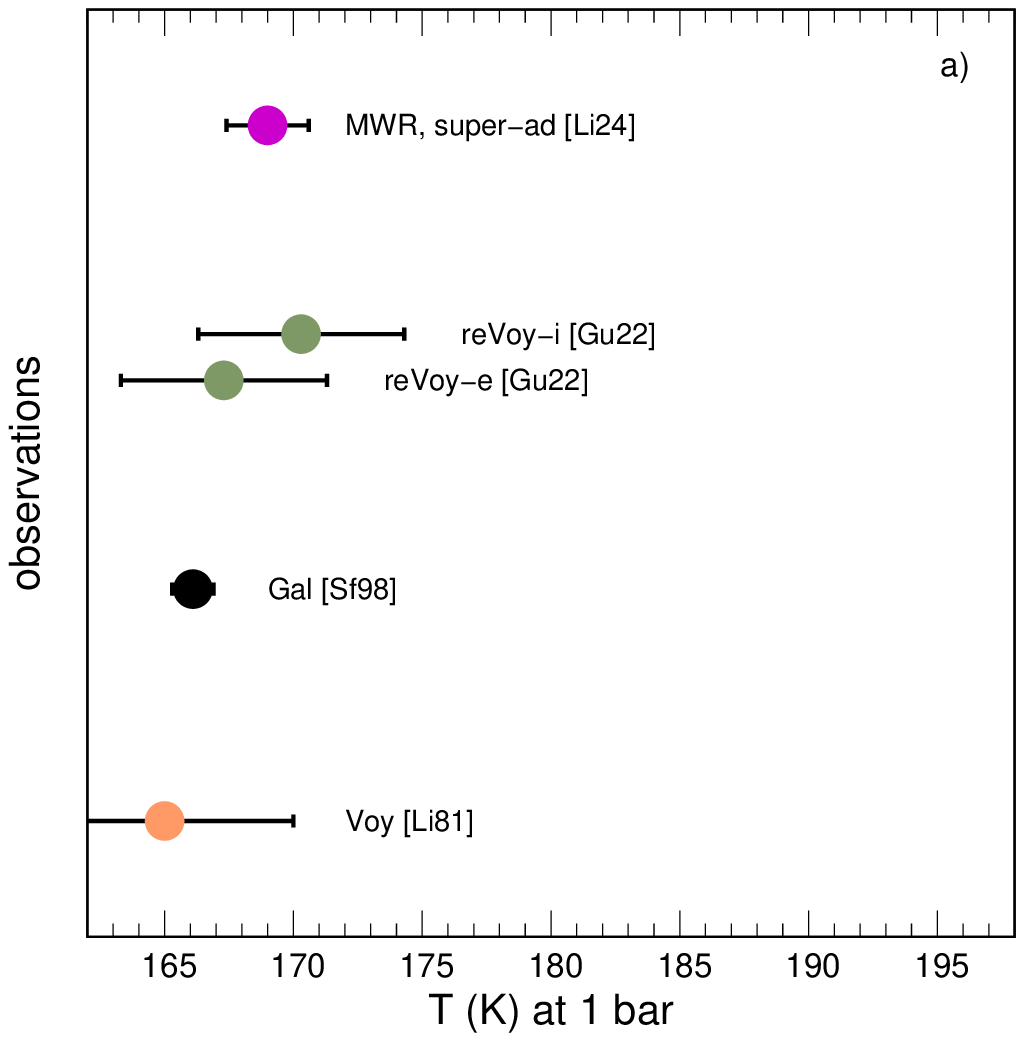}}
\hspace{-2.6cm}
\rotatebox{270}{\includegraphics[width=0.40\textwidth]{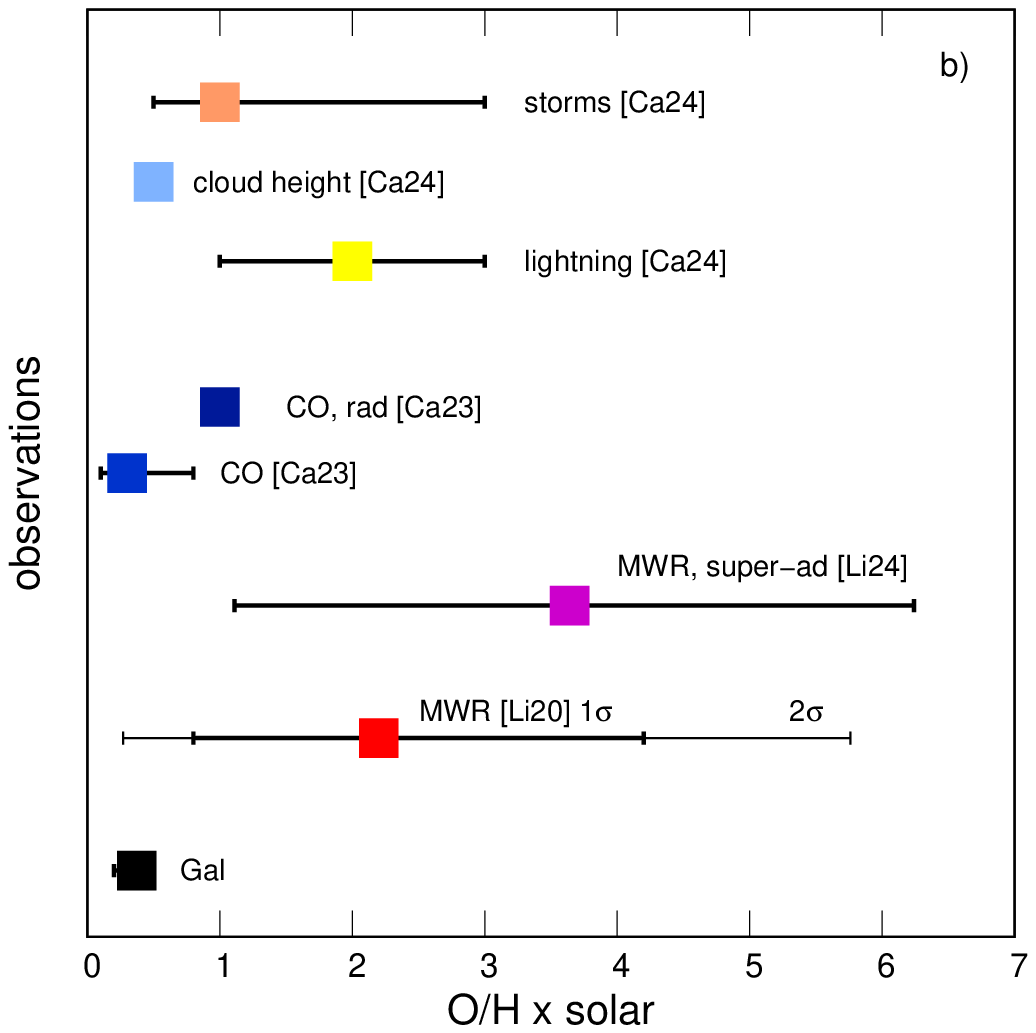}}\\
\noindent
\rotatebox{270}{\includegraphics[width=0.40\textwidth]{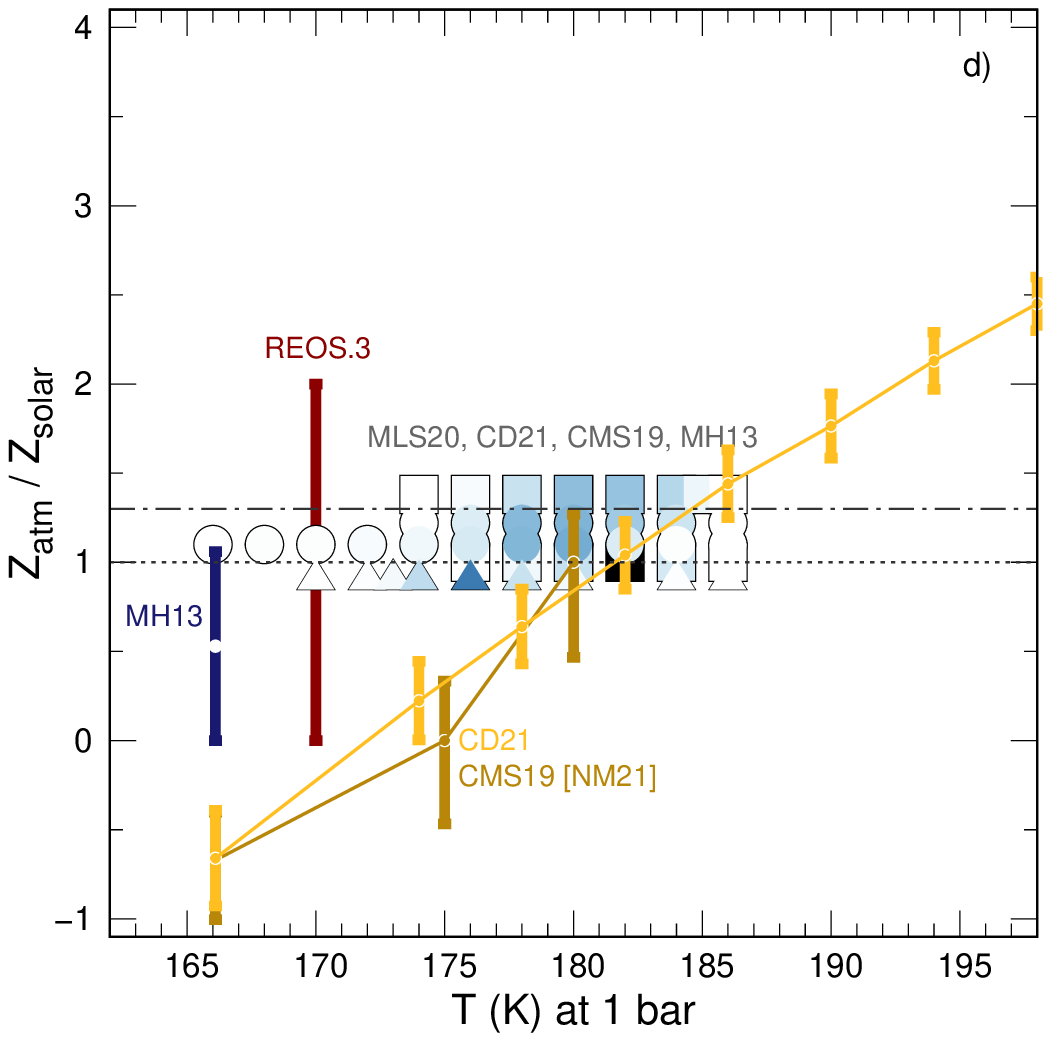}}
\hspace{-2.6cm}
\rotatebox{270}{\includegraphics[width=0.40\textwidth]{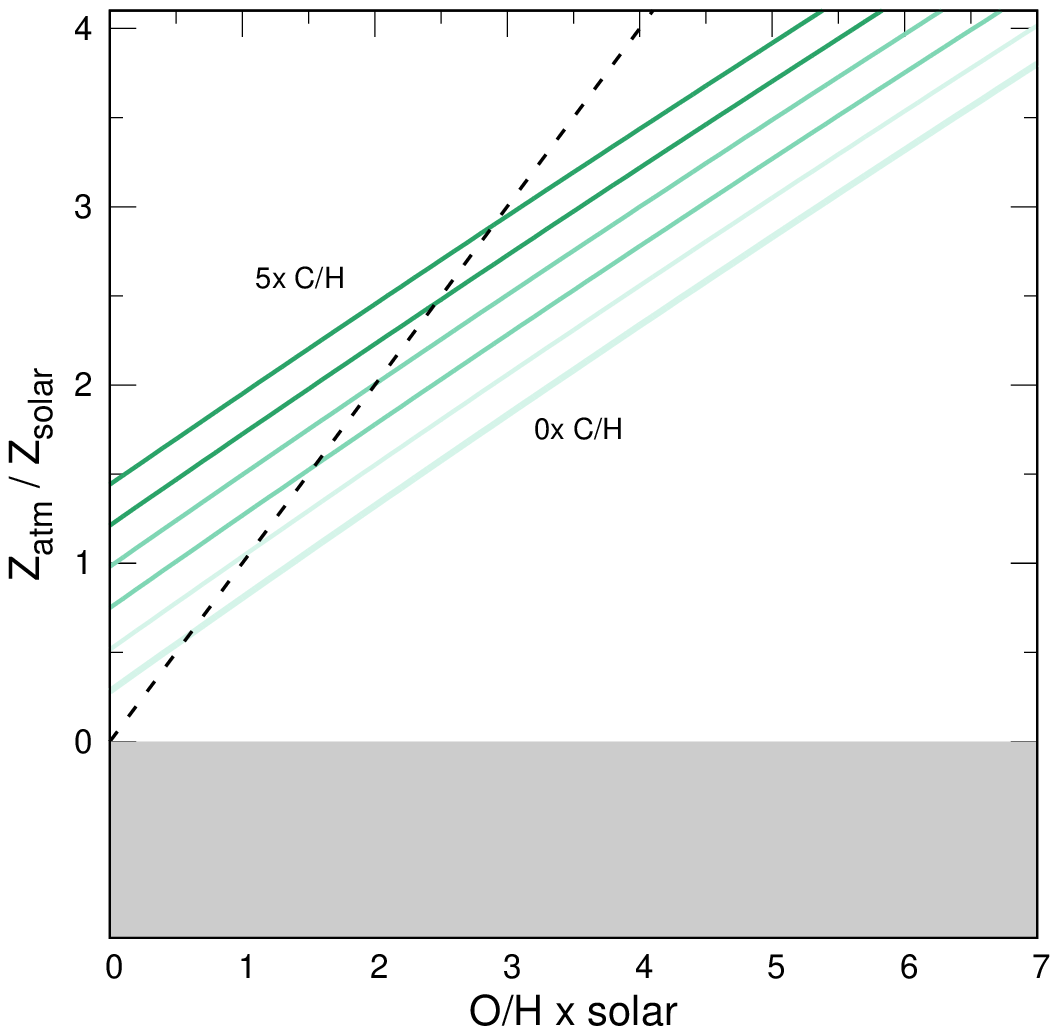}}
\end{minipage}
\caption{\label{fig:OTZ}
$\Tone$--O/H--$\Zatm$ phase space for Jupiter with a scaling of $Z$ by a solar value of  $Z_{\odot}=0.015$. 
Upper left panel (a): observed 1-bar temperatures and the virtual 1-bar temperature of [Li24]; 
upper right panel (b): observationally inferred O/H values scaled by the solar O/H of \citet{Lodders21}; 
lower right panel (c): conversion between O/H and $\Zatm$ with the black dashed line indicating the 1:1 relation as a guide to the eye and with C/H varied from 0$\times$ (light) to 5$\times$ solar (dark) using the solar abundance values of \citep{Lodders21}; lower left panel (d): interior model predictions for the $\Tone$--$\Zatm$ relation by different H/He-EOS, 
REOS.3: [N17], MH13: [MH23], likelihood distributions with MLS20,CMS19,MH13 EOS at 1.0x solar but here slightly 
offset from 1.0: [MB22], same with CMS19 and CD21 EOS at 1.3 x solar but slightly offset from 1.3: [HG23]. 
Intense bluish color indicates high likelihood while pale or white color indicates low likelihood. 
Refs.: Gal=\citet{Wong04} [Gu22]=\citet{Gupta22}, [Li20]=\citet{Li20}, [Li24]=\citet{Li24}, [Ca23]=\citet{Cavalie23},
[Ca24]=\cite{Cavalie24}, [MH24]=\citet{Militzer24}, [HG23]=\citet{Howard23}, [MB22]=\citet{Miguel22}, NM21=\citet{N21}
}
\end{figure*}

The described estimates of O/H in Jupiter's atmosphere from observations are displayed 
in Figure \ref{fig:OTZ}b and scaled to the solar system O/H abundance of \citet{Lodders21} 
of O/H$_{\rm solsys}=6.59\times 10^{-4}$. In particular, for an observed abundance 
$X_{\rm H2O}=2.5^{+2.2}_{-1.6}\times 10^{-3}$ppm \citep{Li20} and the Galileo probe value 
He/H=0.078 we find from 
\[
	X_{\rm H2O} \:=\: \frac{\rm O/H}{\rm H_2/H + He/H + O/H}\:
\]
that O/H$=\:2.2^{+2.0}_{-1.4} \times$ O/H$_{\rm solsys}$, in agreement with \citep{Cavalie24}.

\subsection{Conversion between O/H and Z}

To convert O/H in solar units to Z we use the solar system abundances of \citet{Lodders21} for O,C,N, 
while for the noble gases and the volatiles S,P we use the mean values of the atmospheric abundances as observed
by the Galileo entry probe. Out of these, the largest influence in Z comes from C, so we vary it between 0 and $5\:$x. Furthermore, Z here refers to the mass fraction of heavy elements and not the mass fraction of the molecules they appear in, like water.
The relation between O/H and Z for different methane abundances is displayed in Figure \ref{fig:OTZ}c, 
where $Z$ is scaled by $Z_{\rm solar}=0.015$. With this scaling, O/H = 0x, 1.0x, 2.0x, 3.0 times solar 
corresponds to $Z = (1.0\pm 0.5)\times$, $(1.4\pm 0.6)\times$, $(1.8\pm 0.6)\times$, and $(2.2\pm 0.6)\times Z_{\rm solar}$.

Table \ref{tab:YZobs} lists some estimates for the protosolar metallicity and the protosolar He-abundance with respect to the H-He subsystem. With $\pm 10\%$, the uncertainty in $Z_{\rm solar}$ is  smaller than the  different estimates for individual elements . For instance, the neon abundance of \citet{Lodders21} is 45\% higher than in \citet{AndGrev89} and O/H is 30\% higher than in \citet{Asplund09}. Note that subtracting the mass abundance of neon, which is severely depleted in Jupiter's atmosphere, of 0.0024 from a $Z_{\rm solar}=0.0173$ \citep{Lodders21} yields an effective $Z_{\rm solar}$ of 0.149 for the atmosphere of Jupiter. Throughout this work we assume $Z_{\rm solar}=0.015$. We also use $Y_{\rm proto}=0.270$ with respect to the H-He. We note that this value appears to be too low by 0.005-0.010 compared to the protosolar value inferred from various observations (Table \ref{tab:YZobs}). 

\begin{table}
\centering
Observational estimates of protosolar or solar neighborhood Z and Y\\
\begin{tabular}{llc}
$Z_{\rm solar}$ & $Y_{\rm proto}^{(HHe)}$ & Ref.\\\hline\hline
 0.0213 &  0.275 & AG89\\
 0.0149 &  0.278 & Lo03\\
 0.0142 &  0.274 & AG09\\
 0.0173 &  0.282 & Lo21\\
 0.0140 &  0.280 & NP12, BJA\\
\hline
 0.015  & 0.2700(4)& 
\end{tabular}
\caption{\label{tab:YZobs}Mass fractions of heavy elements and helium abundance sum up to
$1=X+Y+Z_{\rm solar}$ with $Y_{\rm proto}^{(HHe)}=Y/(1-Z_{\rm solar})$.
Last line displays the values used here.
Refs.:[AG89]=\citep{AndGrev89}, [AG09]=\citep{Asplund09}, [BJA]=\citet{BenJaffelAbbes15}
[Lo03]=\citep{Lodders03}, [Lo21]=\citep{Lodders21}, [NP12]=\citet{NievaPrzybilla12} }
\end{table}

\subsection{$Z_{\rm atm}$ from interior models}

\citet{Miguel22} and \citet{HG23} performed extensive Bayesian statistical studies with $\Tone$ priors 
ranging from 135 to 215 K and $\PHe$ ranging from 0.8 to 9 Mbars. For a given $\Zatm$ of respectively $1\times$ and $1.3\times$ solar, \citep{Miguel22} and \citep{Howard23} find that the $\Tone$ posterior distributions peak at around 178-185 K with no solutions in the measured $\Tone$ range of 166--174 K for CMS19 and CD21 EOS.
For $\PHe$, they find a preference in the 1-4 Mbar range. These two studies confirm the result of 
\citet{N21}, who fixed $\PHe$ at 2 Mbar and did not obtain a solution with positive $\Zatm$ at 166.1 K 
for CMS19 EOS while 1x solar $\Zatm$ when $\Tone$ was arbitrarily enhanced to 180 K. Continuing the question of how much $\Tone$ would have to be enhanced, but now with the goal to obtain 2x solar and with CD21 EOS, the answer is 196 K as shown in Figure \ref{fig:OTZ}d. 

With the MH13 EOS \citet{Howard23} also finds a solution for 166 K while with the MLS20 EOS, near 174 K at the upper limit of current observational constraints on $\Tone$.
Using their MH13-EOS, \citet{Militzer22} obtain an optimized Jupiter model (A) with $1\times$ solar $\Zatm$ at $\Tone=166.1$ K that has an extended inhomogeneous, super-adiabatic He-rain region 
between 0.93 and 4.4 Mbars and five layers. A wider parameter range in terms of number and location of layers does not allow for higher atmospheric-Z than this optimized five-layer model A \citep{Militzer24}. 

With REOS.3, up to 2$\times$ solar is possible at an assumed $\Tone$ of 170 K if $\PHe$ is placed rather deep inside at 6 Mbar \citep{N17}. However, $\Zatm$ drops to 1x solar at 4 Mbar and further to 0 for $\PHe < 2\:$ Mbars. 
This behavior demonstrates the importance of stricter constraints on the H/He phase diagram and the $T$-profile across the He-rain region, as the He-poor/He-rich transition zone around $\PHe$ becomes steeper and moves farther out the steeper the $T$-profile is. 

From this compilation one may conclude that Jupiter's atmosphere is at $\sim$$1\times$ solar $\Zatm$ and that MH13 EOS, ML20-EOS, and REOS.3 in the relevant sub-Mbar region come closest to the real H/He-EOS as all other combinations require a higher than observed 1-bar temperature or adiabat-modifications up to 10\%. 
But what if Jupiter's atmosphere is enriched in heavy elements by a factor 2--3 like the heavy
noble gases are, or even more as is possible within the current uncertainty in the atmospheric water abundance?

\section{Methods}\label{sec:meth}

We assume a simple layered structure with six layers. They begin at pressures of 1-bar, 0.1 GPa$=1$ kbar, 2 GPa, $P_{\rm He}$, $P_Z$, and the mass of the compact core $M_{\rm core}$ and are described below.

{Layer No.1 (L1)} extends from the 1-bar level in the atmosphere to the top of 
the OSL at 1 kbar. It is adiabatic, of $Z$-level $Z_1$ that is referred to as $\Zatm$, 
and constant He mass abundance $Y_1=0.238$ consistent with the Galileo entry probe measurement 
of $Y_{\rm Gal}=0.238\pm 0.005$. The 1-bar temperature is set to 170 K.
In the absence of an additional inverted Z-gradient, $Z_1=Z_2=Z_3=Z_4$. In the presence of an additional Z-gradient $\Delta Z<0$, we fix $\Zatm$ to, e.g., $3\times$ solar while $\Delta Z$ is adjusted to ensure $Z_3=\Zatm + \Delta Z$.

{Layer No.2 (L2)} is the OSL. We assume that there is a physical reason why the OSL exists
without specifying it here, although we believe a radiative window due to low
opacity is the only viable explanation. The OSL is assumed to have an inverted He-gradient $\Delta Y<0$ and in some models an additional inverted Z-gradient $\Delta Z<0$. Both gradients are assumed to be linear with pressure. 
To stabilize the compositional gradient against convection we assume that Ledoux-stability does that job. Here, we pick a favorable value of the density ratio $R_{\rho}^{-1}=0.9$ $(R_{\rho}=1.11)$
where 
\begin{equation} \label{eq:R0inv}
R_{\rho}^{-1} = \frac{\alpha_{\mu}}{\alpha_T}\:\frac{\nabla_{\mu}}{\nabla_{T}-\nabla_{ad}}
\end{equation}
and $\alpha_T=d\ln \rho/d\ln T$ is the thermal expansion coefficient and $\alpha_{\mu}=d\ln \rho/d\ln \mu$. 
Values $1<R_{\rho}^{-1}<R_{\rm crit}^{-1}$ indicate the regime of double diffusive convection with 
stabilizing compositional and de-stabilizing super-adiabatic thermal gradient, which can further 
be subdivided into layered  and oscillatory double-diffusive convection \citep{Mirouh12,Tulekeyev24}. 
If the compositional gradient is de-stabilizing, as in our case, and the temperature gradient stabilizing, values $1<R_{\rho}<R_{\rm crit}$ indicate another regime of double diffusive convection \citep{Brown13}, which can occur in the form of fingering double diffusive convection. This process has been suggested to occur in red giant stars as a mechanism to enhance the mixing between a ${}^3$He-rich layer and an adjacent deeper layer where ${}^3$He can burn to a mix of ${}^4$He and protons, which has a lower mean molecular weight, although the mixing by fingering double diffusive convection still seems to be too inefficient to explain the observed ${}^3$He depletion in the atmospheres of RGB stars \citep{Langer10,Wachlin14}.

To Ledoux-stabilize the OSL, we adjust the sub-adiabatic temperature gradient $\nabla_T$ to achieve a value $R_{\rho}^{-1}=0.9$. Values greater than 1 would cause a transition to overturning convection, 
while lower values would imply a stronger sub-adiabaticity and thus colder interiors. With our
choice of $R_{\rho}\sim 1.1$ we also ensure to not pass the threshold to the purely diffusive regime,
which occurs for $R_{\rho} > R_{crit}$ with $R_{\rm crit} = 1/\tau$ \citep{Rosenblum11}. $R_{\rm crit}$ 
adopts values of 2--3 across our OSL, and $\tau$ is the particle diffusivity, here of He in the He-H system.

\begin{figure}
\centering
\rotatebox{270}{\includegraphics[width=0.4\textwidth]{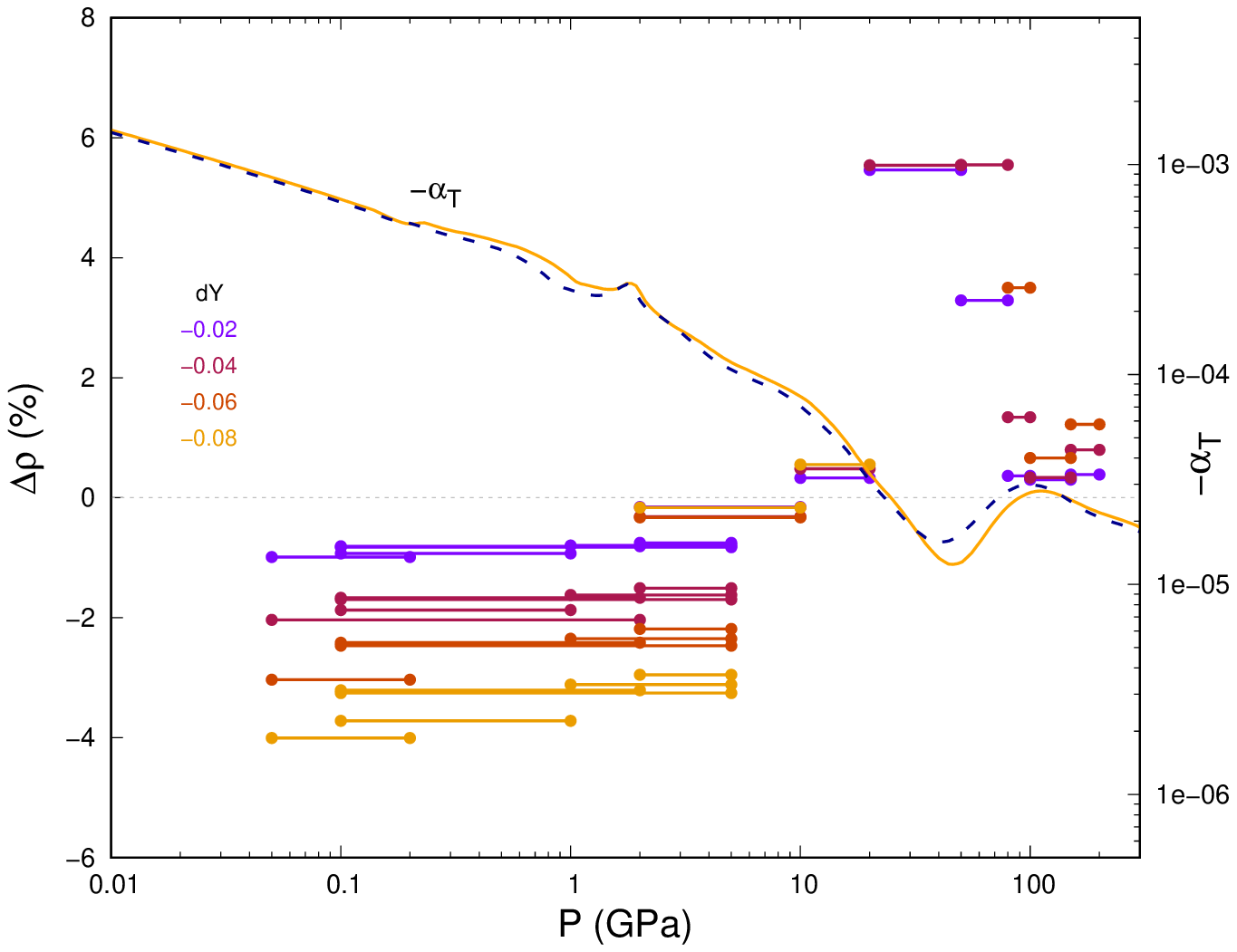}}
\caption{\label{fig:optimOSL}
Relative density differences (in \%) at Jupiter's 2 Mbar level between H/He $P$--$T$ profiles with and without an OSL for different de-stabilizing He-depletion levels  (color code). The bottom axis shows the pressure-interval where the OSL is assumed to occur. The decrease in density is strongest if the OSL is placed at about 1 kbar (0.1 GPa), where the thermal expansion coefficient $\alpha_T$ (black line for protosolar Y=0.27, orange line for Y=0.16) is high) while if placed at the dip of $\alpha_T$, the density at 2 Mbar increases dramatically.}
\end{figure}

The OSL is assumed to extend between 1 and 20 kbar (which is 0.1--2 GPa, or 0.99--0.975 RJ, or 1300--2300 K). 
This choice is motivated by a broad-band optimization in a sense that such location is found  to lead 
to higher-$\Zatm$ values than others. The optimization is not fine-tuned in a sense that we did not 
search for the global optimum location. In particular, a placement in the 10--100 GPa region was found 
to cause the opposite trend of a much reduced resulting $\Zatm$. That our non-exhaustive optimization 
favored this $P$-range of 1-20 kbar does not exlude slightly higher or lower, or more or less extended 
locations. It is important, however, that the OSL is placed farther out than the 10 GPa level. 
The decrease in the resulting $\Zatm$ if the OSL is placed at intermediate pressures of 20-80 GPa is due to the minimum in  $\alpha_T$ there, see Figure \ref{fig:optimOSL}. For a given $R_{\rho}$, a lower $\alpha_T$ in Eq.\ref{eq:R0inv} requires a stronger sub-adiabaticity, which makes the interior colder and thus denser. 

Figure \ref{fig:optimOSL} shows the relative change in density between H/He $P$--$T$ profiles with and without an OSL, here at a reference pressure of 2 Mbar. In Jupiter, the 2 Mbar pressure level
occurs at 0.81 $\RJ$. There, both gravitational harmonics $J_2$ and $J_4$ that we use to constrain the $Z$-profile are highly sensitive to density as parameterized by their contribution functions \citep{HelledMN22}. A decrease in density up to 4\% percent is obtained if the OSL is placed higher up in the planet than the 10 GPa level, and of course the density response increases with increasing $\Delta Y$. However, the response in $\Zatm$ can differ from from the raw local density response because the sensitivity ranges of $J_4$ and $J_2$ are quite large. For instance we found a small increase in the resulting $\Zatm$ if the OSL was placed at around 1 Mbar although according to
Figure \ref{fig:optimOSL}, the density of a H/He profile would increase deeper down. The increased density deeper down leads to a reduction in the deep-Z level, which reduces $|J_4|$ and thereby reduces the need to keep $\Zatm$ low. This coupling allows for a slight enhancement in $\Zatm$ when the OSL is placed at the 1 Mbar level, but with only $(0.25\pm 0.30)\times$ the influence of an OSL, on $\Zatm$, we found this effect to be too small to be useful. 
With a placement farther out, at 0.1--2 GPa, the density at 2 Mbar is reduced by $\sim 3.5 \%$ at the 2 Mbar level.

Layer No.3 (L3) is the most He-depleted layer and of He mass fraction $Y_3=Y_{\rm Gal}+\Delta Y$. 
The He-depletion level is assumed to result from H/He-demixing and would ideally be predicted by 
the H/He-phase diagram. However, those have large uncertainties; in addition, the He-depletion level 
predicted by a given H/He-phase diagram depends on the EOS used \citep{N15,N24}. 
As none of the available phase diagrams predicts right-away the observed He-depletion, it has become 
common to shift the phase diagram in temperature or pressure 
\citep{FH03,N15,Puestow16,Mankovich16,MF20,Howard24}. Therefore, a stronger-than-observed He-depletion 
level does not pose an inconsistency with the H/He phase diagram, just the required shift may be different. 
L3 begins at the bottom of the OSL and extends into the He-rain region up to a pressure $\PHe$. 

For simplicity, we do not model the He-rain region explicitly but represent it by a sharp increase 
in He-abundance  at $\PHe$. In real Jupiter, the He-rain region may extend over several Mbars down
to 4 Mbars \citep{N15} or deeper, corresponding to $\sim 10\%$  in radius. Even larger extensions are 
sometimes suggested from superposition of Jupiter adiabats with demixing curves at constant 
He-concentration \citep{Brygoo21,Chang23}. 

\begin{figure}
\centering
\rotatebox{0}{\includegraphics[width=0.5\textwidth]{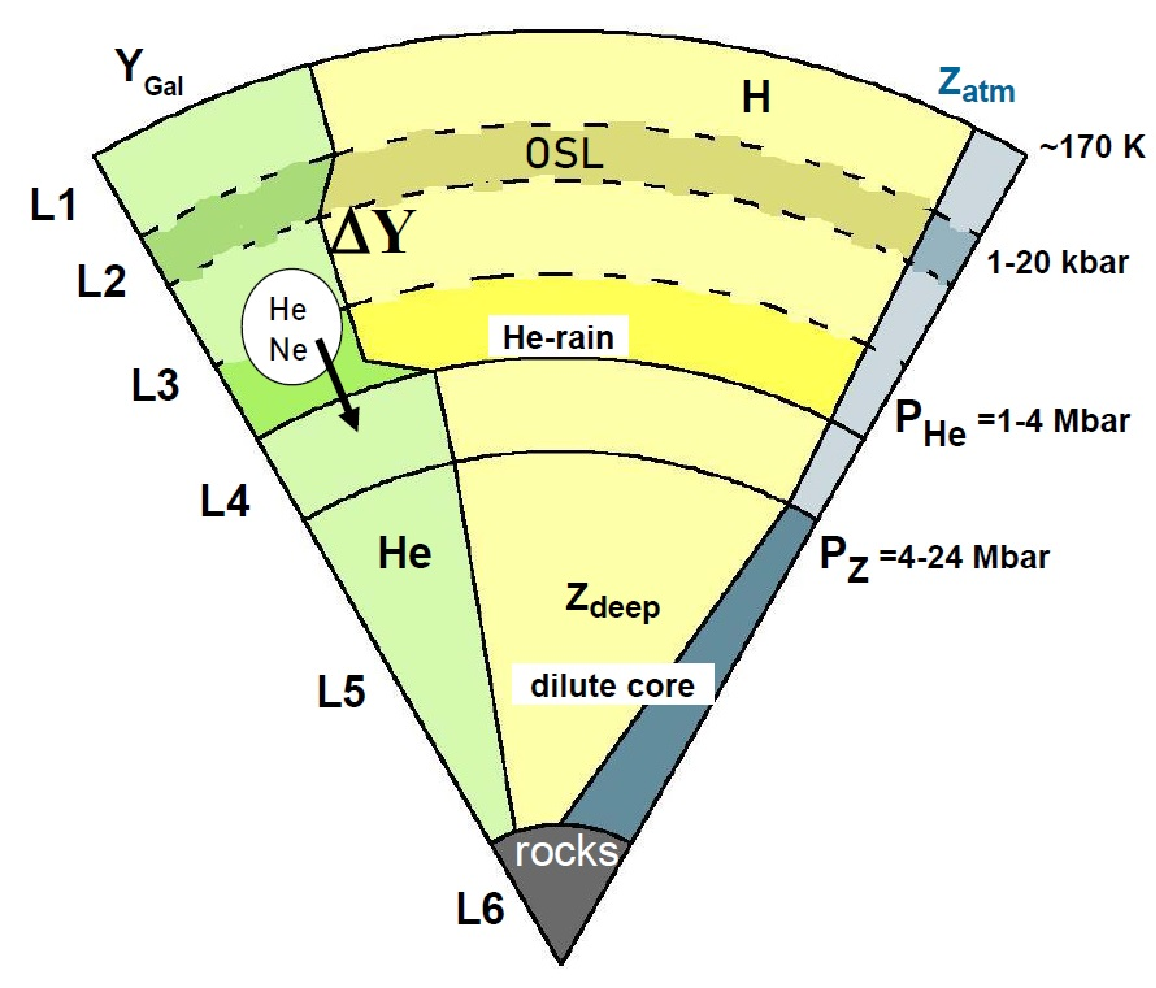}}
\caption{\label{fig:torteJ}
Illustration of the Jupiter layered structure model with inverted He-gradient $\Delta Y<0$ across the OSL. The heavy element mass abundance $\Zdeep$ underneath $\PZ$ is shown with a gradient, although most models are computed with constant-$\Zdeep$ and only few models with a Gaussian gradient. $\Zatm$ extends from the 1~bar level either down to $\PZ$ as illustrated, or in models with additional inverted Z-gradient only to the top of the OSL. The He-gradient in the He-rain zone is illustrated here to begin shallow then be linear, while in the model computations we make a sharp transition at $\PHe$. 
The radial positions of the layers are not to scale. 
}
\end{figure}

The extent of the He-rain zone and the He-gradient therein depend not only on the H/He phase diagram but also on 
the P-T profile, which in turn depends on EOS and the assumed heat transport 
mechanism \citep{FH03,N15,Mankovich16}. The LHR0911 phase diagram shifted in temperature 
leads to an about linear increase in He-abundance across the He-rain region \citep{N15,Mankovich16}. 
However, the LHR0911 phase diagram predicts H/He demixing, which is related to the metallization of H, to begin at 1 Mbar while experiments on fluid H find the phase transition to metallic H to occur 
at 1.4 Mbar \citep{Weir96}.  A pressure shift of the LHR0911 phase diagram by 0.4 Mbar so that demixing in Jupiter would begin at 1.4 Mbar was found to lead to a very shallow He-gradient, 
due to the planetary P-T profile in that case running parallel to the phase boundary \citep{N15}. 
Such behavior implies that the He-poor region, i.e.~our L3, could extend far into the He-rain region 
before the He-abundance begins to rise. This possibility is illustrated in Figure \ref{fig:torteJ}. 
The simplified sharp transition adopted in this work and its placement somewhere within the He-rain region is meant to be representative of a gradual tradition that however may begin very shallow. Given current uncertainty in the H/He phase diagram, we simply vary $\PHe$ between values of 1 and 4 Mbar. 
 
{Layer No.4 (L4)} begins at a pressure $\PHe$ and extends to a pressure $P_{Z}>\PHe$, where $\PHe$ marks the He-poor/He-rich transition associated with the He-rain region while $\PZ$ marks a the onset of a dilute core with Z-level $Z_{\rm deep}$. Starting with $\PHe$, the H/He ratio is kept constant at the He-enriched level that is required to meet a mean H/He abundance of $Y_{\rm mean}=0.2700(4)$. Higher assumed $Y_{\rm mean}$ values, as Table \ref{tab:YZobs} shows may be a better representation of $Y_{\rm proto}$ and would lead to lower $\Zdeep$ values. L4 has $Z_4=Z_3$. In classical three-layer models, $\PHe=\PZ$ and this layer drops out.

{With Layer No.5 (L5)} begins the region where the heavy element abundance can differ from that in layers 1--4, or in the case of an inverted Z-gradient, from that in layers 3--4, $Z_{\rm deep}=: Z_5$. This layer extends from $P_Z$, which we vary between 4 and 24 Mbars, all the way down to a possible compact rock core. Models without OSL ($\Delta Y=0$) typically have $\Zdeep \gg \Zatm$ with $\Zdeep$ 5--15$\times$ solar. This central region is thus rather dilute. 
We either assume a constant-Z level for L5 or a half-Gaussian profile. In the latter case, the maximum near the core-mantle boundary is the $Z_5$-level.

Both $Z_3$ and $Z_5$ are used to adjust $J_2$ and $J_4$. In models with $\Delta Z=0$, all Z-levels reduce to the two constant levels $\Zatm$ and a $\Zdeep$. In models with inverted Z-gradient, there is $Z_3$ in addition to $\Zatm$ and $\Zdeep$.

{Layer No.6 (L6)} is a compact rock core. The core mass is used to ensure the boundary condition $m(r=0)=0$ for given $\MJ$, $\RJ$.

For H/He, we use the CD21-EOS \citep{CD21} while for heavy elements, we use $\rho(P,T)$ of different volatile EOSs. These are the extrapolated Ice-EOS of \citet{HM89} at low pressures of 1-1000 bars and 
H2O-REOS at $P\geq 1\:$kbar. The heavy elements are neglected in the computation of the entropy because their contribution to entropy is small for $Z\lesssim 0.2$ \citep{Baraffe08}. For the rocky core we use
the $P$-$\rho$ relation for rocks by \citet{HM89}.

\section{Results}\label{sec:res}

\subsection{Z-levels of models with inverted He-gradient}

\begin{figure*}
\centering
\hspace*{-1cm}
\rotatebox{270}{\includegraphics[width=0.70\textwidth]{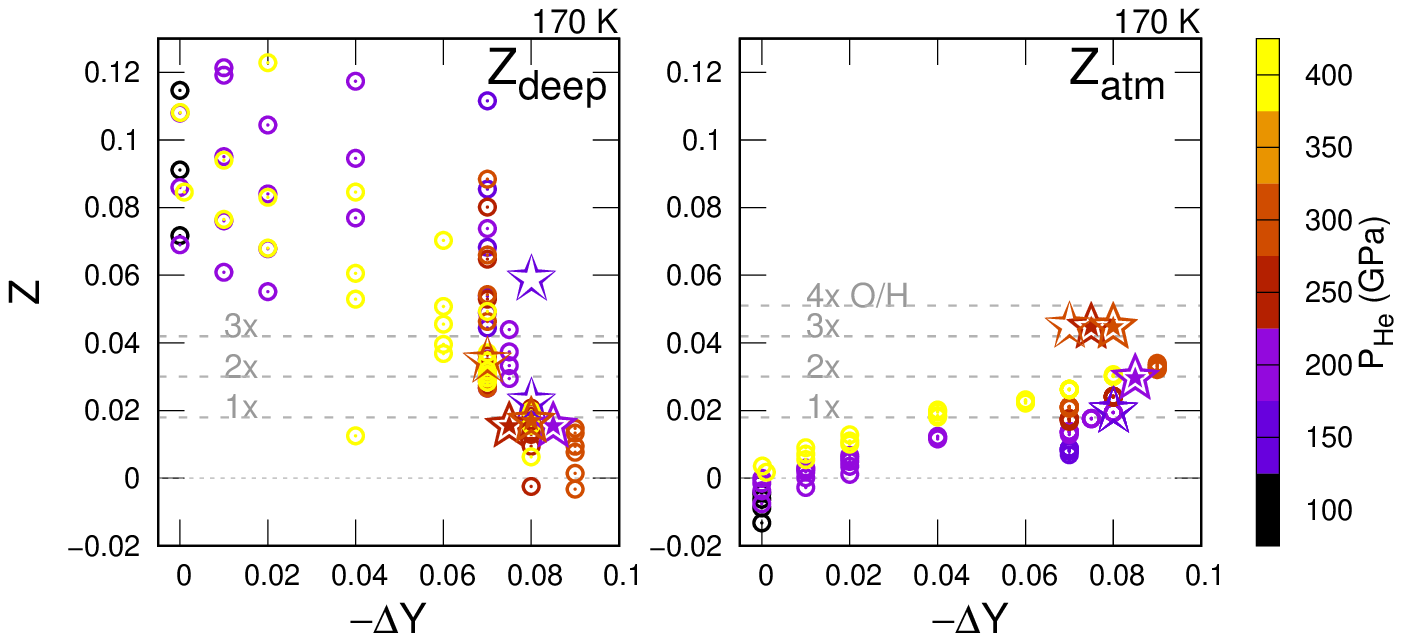}}
\caption{\label{fig:ZZ_dYdZ}
Atmospheric Z (right panel) and deep interior Z (left panel) of Jupiter models with inverted He-gradient $\Delta Y\leq 0$ (x-axis) across the OSL and different values of $\PHe$ (color code) as well as of $P_Z$. Horizontal grey dashed lines indicated levels of enrichment over a solar $Z_{ \odot}=0.015$. Star symbols indicate models with additional inverted Z-gradient, among those, open stars show models where $Z_3\lesssim 1\times$ solar while pattern-filled stars show models with $Z_3 \gtrsim 1\times$ solar.}
\end{figure*}

Figure \ref{fig:ZZ_dYdZ} shows a compilation of our models with an inverted He-gradient (dots) and of 
models with an additional Z-gradient (stars) in terms of  $\Zatm$ and the $\Zdeep$ level.  The models exhibit several kinds of behavior that is well-known from $\Delta Y=0$ models.
First ($i$), with a deeper He-poor/He-rich transition pressure $\PHe$, $\Zatm$ increases. This behavior has been demonstrated in print mostly for classical three-layer models where $\PHe=\PZ$ \citep{N12,Militzer24}. 

Second $(ii)$, for moderate $|\Delta Y| \lesssim 0.04$, the models have $\Zdeep \gg 1\times$ solar and $\Zdeep > \Zatm$, i.e., a heavy-element enriched deep interior that is more enriched than the molecular region above the He-rain zone  \citep{N08,N12,Wahl17J,N17,N21,Miguel22,Militzer24,Howard23}. That spread in the $\Zdeep$ values is due to the choice of $\PZ$. One may call the heavy element-enriched interior a dilute core. Furthermore, these models also have a low $\Zatm<1\times$ solar (O/H$<0.5\times$ solar) unless $\PHe$ is placed deep, at 4 Mbars, in which case $1\times$ solar (yellow dots) is obtained.

The lowest order gravitational harmonic $J_2$ is more sensitive at Mbar pressures than $J_4$ is, with a sensitivity  that reaches down to about 0.6 $\RJ$ where $P$ about 10 Mbar \citep{HelledMN22}. Placing $\PZ$ deep reduces the density above  $\PZ$ (farther out), which tends to reduce $J_2$. This reduction in $J_2$ can be compensated by an enhancement
in density below $\PZ$ (deeper inside). Therefore third $(iii)$, $\Zdeep$ rises with $\PZ$.  This behavior has been 
demonstrated for classical three-layer models \citep{N12} but also for four-layer models \citep{Militzer24}.

The models exhibit further characteristic behavior that is related to $\Delta Y< 0$. Fourth ($iv$), --as expected and intended-- the possible $\Zatm$ level rises with increasing He-depletion at depth.

Fifth $(v)$, in models with high $|\Delta Y| \gtrsim 0.07$ we see the new behavior of a maximum $\Zdeep$ 
that quickly drops while $\Zatm$ keeps on rising. This can be understood as follows. The lower the He-abundance in L3, the more He must be present in the deeper layers L4--L5 to conserve a bulk He-abundance $Y_{\rm mean}$. A higher He-abundance along a H/He adiabat implies higher densities. This effect applies to pressures $P> \PHe$, ie to both L4 and L5.
Furthermore, a lower He-abundance in L3 requires a larger sub-adiabaticity in L2, and therefore the interior throughout L3---L5 becomes colder. The thermal EOSs of H and He are still sensitive to temperature at a few Mbars up to $\sim$ 10 Mbars \citep{N08}. Therefore a colder H/He-adiabat implies a denser H/He-adiabat through L3 and into L4, and depending on $\PZ$ also into L5. 
Both effects (more He, colder) tend to enhance the density where $J_2$ is still sensitive. To conserve $J_2$ then requires a lower $\Zdeep$ value. Eventually, $\Zdeep < \Zatm$ occurs. 

Sixth ($vi$), contrary to the characteristic behavior $(iii)$, placing $\PZ$ deeper at high $|\Delta Y|$ would only increase the density deep down and require even lower $\Zdeep$ values. An imposed lower limit $\Zdeep \geq 0$  implies there is an upper limit on $\PZ$. It moves toward $\PHe$ with increasing $|\Delta Y|$.
When $-\Delta Y$ is increased beyond $\sim 0.09$, the $\PZ$ window closes. 
The fall of $\Zdeep$ imposes an upper limit on $|\Delta Y|$ of $\sim 0.09$, which poses an upper limit on the $\Zatm$-values that can be reached with these kind of models. With an inverted He-gradient only and CD21-EOS, $\Zatm \sim 2\times$ solar can be reached, which is an increment by 0.03 compared to the non-OSL case.

There is a sweet spot where $\Zatm$ passes the 1$\times$ solar threshold while $\Zdeep \geq \Zatm$.
In this case, adiabatic models with a largely homogeneous-$Z$ at 1$\times$ solar occur. 
However, small compositional differences $\Zdeep \gtrsim \Zatm$ may not be stable against double diffusive mixing \citep{Tulekeyev24}, and therefore models with $\Zdeep \gg \Zatm$ may be preferred. A factor of five or more in compositional difference can be achieved by placing $\PZ$ deep, within the inner 0.2--0.3 $\MJ$.
Such models are on the branch  where deeper $\PZ$ leads to higher $\Zdeep$ (characteristic behavior $iii$). Interior $Z(m)$ profiles of such models are shown in Figure \ref{fig:Zm_1x} and will be discussed in Section \ref{sec:Zm}.

To reach beyond $2\times$ solar in $\Zatm$ can be achieved by allowing for an inverted Z-gradient 
in addition to the inverted He-gradient. With this set-up, we can find models up to 3x solar. While 
the region above the OSL comprises only  1.4$\times 10^{-4}$ M$_{\rm J}$ and extends over only 0.01$\RJ$, 
which is about a third of the depth of the zonal winds, it has an effect on the low-order gravitational 
harmonics $J_2$ and $J_4$ and that effect is much larger than that from the winds. The effect occurs indirectly 
through the sub-adiabaticity. 
An additional Z-gradient requires a stronger sub-adiabaticity and thus leads to a cooler adiabat
deeper down where the low-order $J_n$ are more sensitive. Through this effect, higher $\Zatm$ values in L1 
lead to lower Z-values in L3.

\subsection{Interior Z(m)-profiles and the origin of Jupiter}\label{sec:Zm}

\begin{figure}
\centering
\rotatebox{270}{\includegraphics[width=0.50\textwidth]{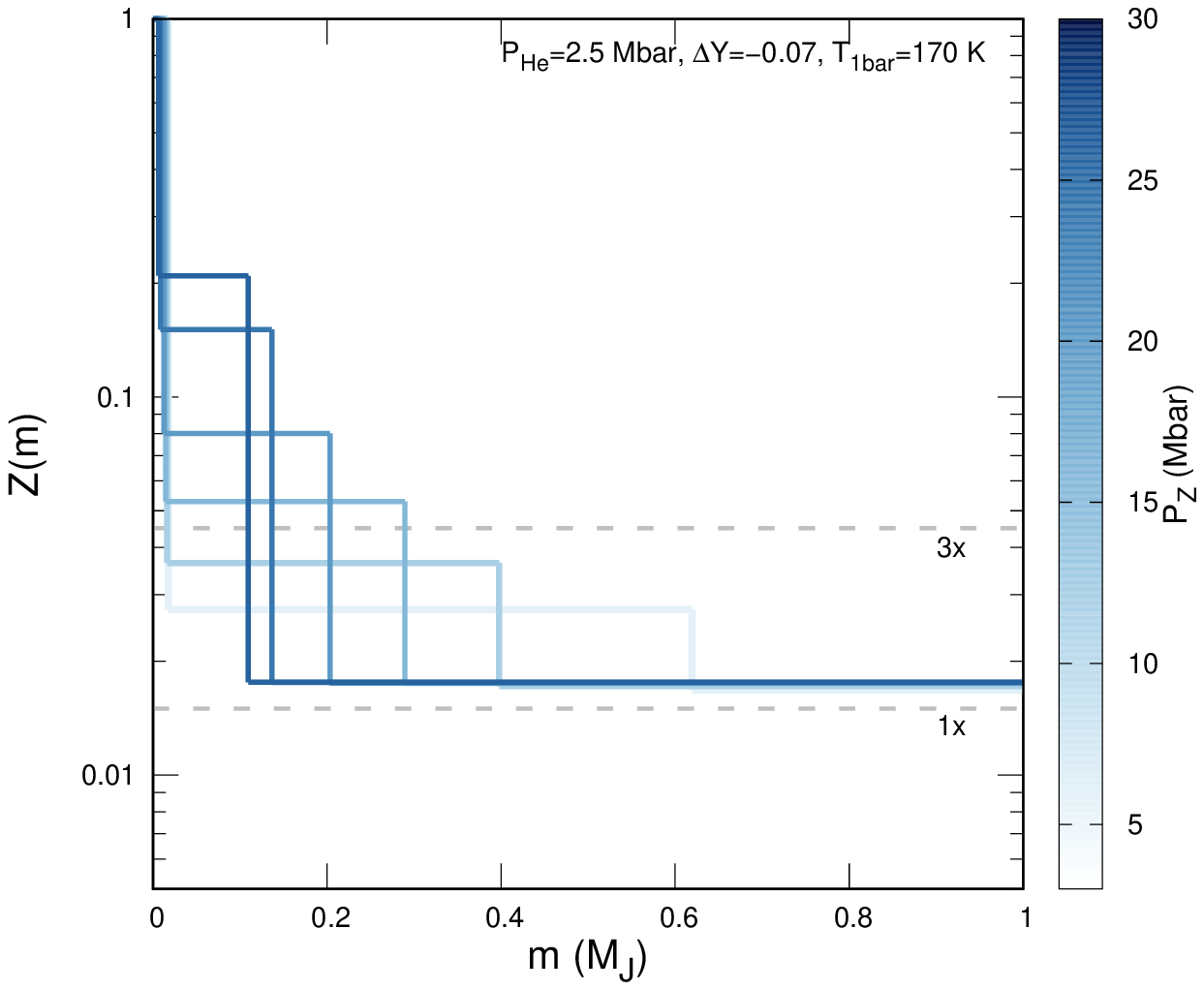}}
\caption{\label{fig:Zm_1x}
Preferred Jupiter models with 1x solar in the atmosphere. As input, these models have an 
inverted He-gradient $dY=-0.07$, $\PHe=2.5\:$ Mbar, $\Tone=170\:$K, and $P_Z$ as given by the color code. 
A dilute core is possible with a substantial (factor 5) increase in the heavy-elements, in which case
this strongly enriched region is confined to the very deep interior (0.2$\:\MJ$).
}
\end{figure}

\begin{figure}
\centering
\rotatebox{270}{\includegraphics[width=0.50\textwidth]{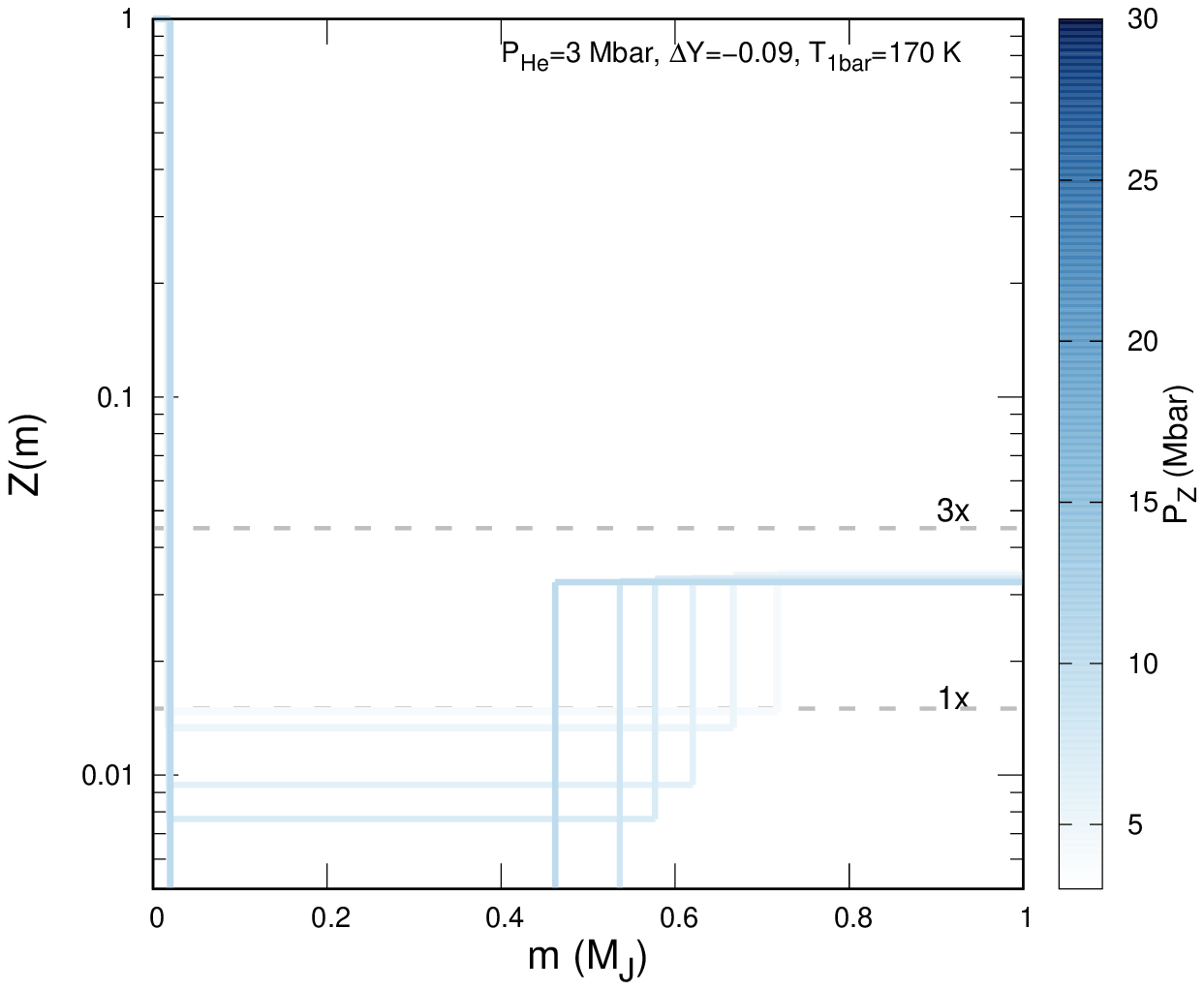}}
\caption{\label{fig:Zm_2x}
Same as Figure \ref{fig:Zm_1x} but for selected models with $2\times$ solar $Z$ in the atmosphere.
As input, the models have an inverted He-gradient ($dY=-0.09$), $\PHe=3\:$Mbar, and $\Tone=170\:$K, 
and $P_Z$ as given by the color code. 
}
\end{figure}

\begin{figure}
\centering
\rotatebox{270}{\includegraphics[width=0.50\textwidth]{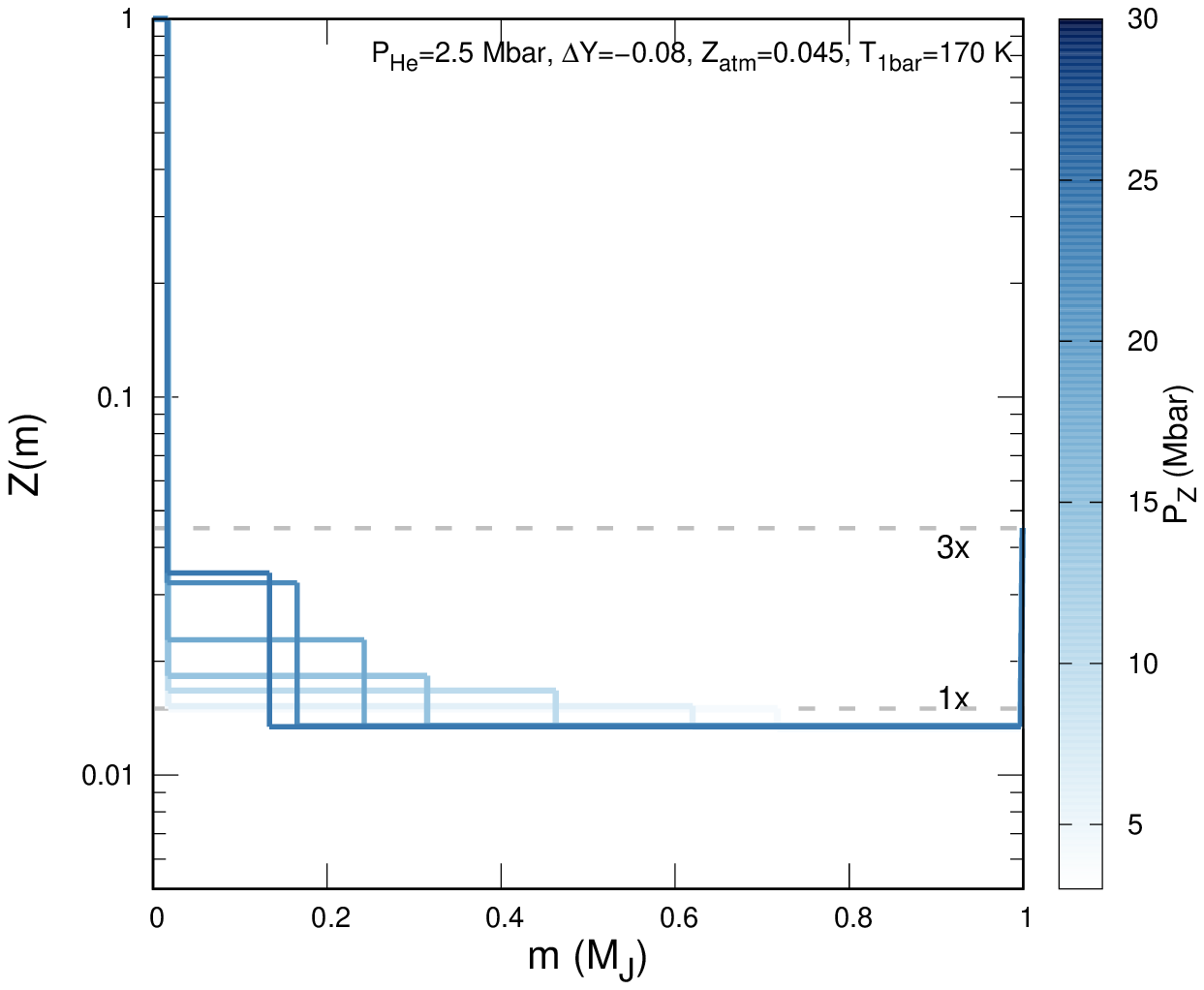}}
\caption{\label{fig:Zm_3x}
Same as Figure \ref{fig:Zm_1x} but for selected models of $3\times$ solar $Z$ in the atmosphere.
As input, the models have an inverted He-gradient ($dY=-0.08$), an inverted Z-gradient, $\PHe=2.5\:$Mbar, 
$\Tone=170\:$K, and $P_Z$ as given by the color code. 
}
\end{figure}

Interior $Z(m)$ profiles are important for a comparison to predictions from planet formation models, 
and also to probe the stability of the Z-gradient against microscopic instabilities. 
We present $Z(m)$ profiles of selected models with 1$\times$, 2$\times$, 3$\times$ solar $\Zatm$ 
in Figures \ref{fig:Zm_1x}--\ref{fig:Zm_3x}. 

Some models with $\Zatm \sim 1\times$ solar (Figure \ref{fig:Zm_1x}) and $\Zdeep \gtrsim \Zatm$ have only 
a small difference between $\Zdeep$ and $\Zatm$. These models may not be stable against convective or semi-convective mixing.
\citet{Tulekeyev24} found that a substantial density gradient is needed to avoid a transition to layered
double diffusive convection, in which case small layers will sooner or later begin to merge and leave behind a
homogenized large layer. However, one must caution that these results are based on simulation cells of limited size
that may not be directly applicable to the large-scale behavior and geological time-scales. Nevertheless, if one
wishes to maximize the Z-gradient in this set of models, $\PZ$ would have to be with $\gtrsim 20$ Mbar, rather 
deep. This confines the dilute core to the innermost 0.2--0.3 $\MJ$, or 0.42--0.5 $\RJ$.

Such models in Figure \ref{fig:Zm_1x} are consistent with predictions from Jupiter formation models. 
\citet{Helled22} show that the accretion of high-Z material is limited to the intermediate phase-2 of giant planet formation, 
and that the Z-enrichment of the accreted material decreases with time and radius so that in the forming Jupiter, 
a gradual Z-distribution emerges. After the run-away gas accretion phase-3, the high-Z region will be confined to the 
innermost 30\% of Jupiter's mass (50\% in present radius), while the homogeneous-Z enrichment of the massive envelope 
will depend on the solids surface density. These Jupiter interior models are also consistent with the proposal that
the phase-2, where both gas and heavy elements are accreted, may last several Myrs so that the heavy element-rich core 
can grow to a substantial size of $\sim 100 \ME$ (0.3$\MJ$) until runaway gas accretions sets in \citep{Helled23}. 
Whether or not the initial central compositional gradient will be subject to semi-convective mixing \citet{Tulekeyev24} 
depends on the initial temperature profile, which is quite uncertain.

Models with high $-\Delta Y\sim$0.09 and $\PZ$ close to $\PHe$ yield a $\Zatm$ of up to 2x solar (Figure \ref{fig:Zm_2x}).
They have a rather narrow transition zone of a few Mbars where the abundances of both He an $Z$ change 
and where the EOS is still sensitive to temperature. While here we restrict ourselves to adiabatic 
He-rain zones, this He-gradient zone may be superadiabatic, although thermal evolution models favor only small 
superadiabaticity \citep{N15,MF20}. A superadiabatic He-rain zone would reduce the density underneath and may 
lift $\Zdeep$, which in this class of models is sub-solar, up to the level of $\Zatm$. Models 
with a super-adiabatic He-rain zone, although not explicitly constructed here, would have a homogeneous-$Z$ 
underneath the OSL and a possible but not mandatory small compact core of $\sim 5\ME$. Such models are consistent 
with the traditional view of rapid gaseous envelope accretion that happens when the accreted gas-component 
surpasses the initial compact core mass \citep{Pollack96}, which may be lower than the commonly found $10 \ME$
depending on the pollution of the accreted gas component \citep{Hori11}. 

An additional Z-gradient requires a stronger sub-adiabaticity to stabilize the OSL against convection.
In the colder interior, a lower amount of heavy elements is needed to achieve the same density as in the case without the additional Z-gradient. Both $\Zdeep$ and the size of the dilute core shrink, leaving behind a largely
1$\times$ solar envelope under a 3$\times$ solar atmosphere (Figure \ref{fig:Zm_3x}). Such an interior is consistent with a short phase-2 in the classical core-accretion scenario of planet formation. 
Such an interior is not consistent with the delayed-phase-2 planet formation model of \citet{Helled23},
which has an extended $\sim 100\:\ME$ dilute core, unless the gas accreted by run-away was sub-solar, 
and if mixing over time has eroded the initial Z-gradient just to enrich the failed-gas-giant 
model homogeneously to about 1$\times$ solar.

\subsection{Core mass and $J_6$}

\begin{figure}
\centering
\rotatebox{270}{\includegraphics[width=0.50\textwidth]{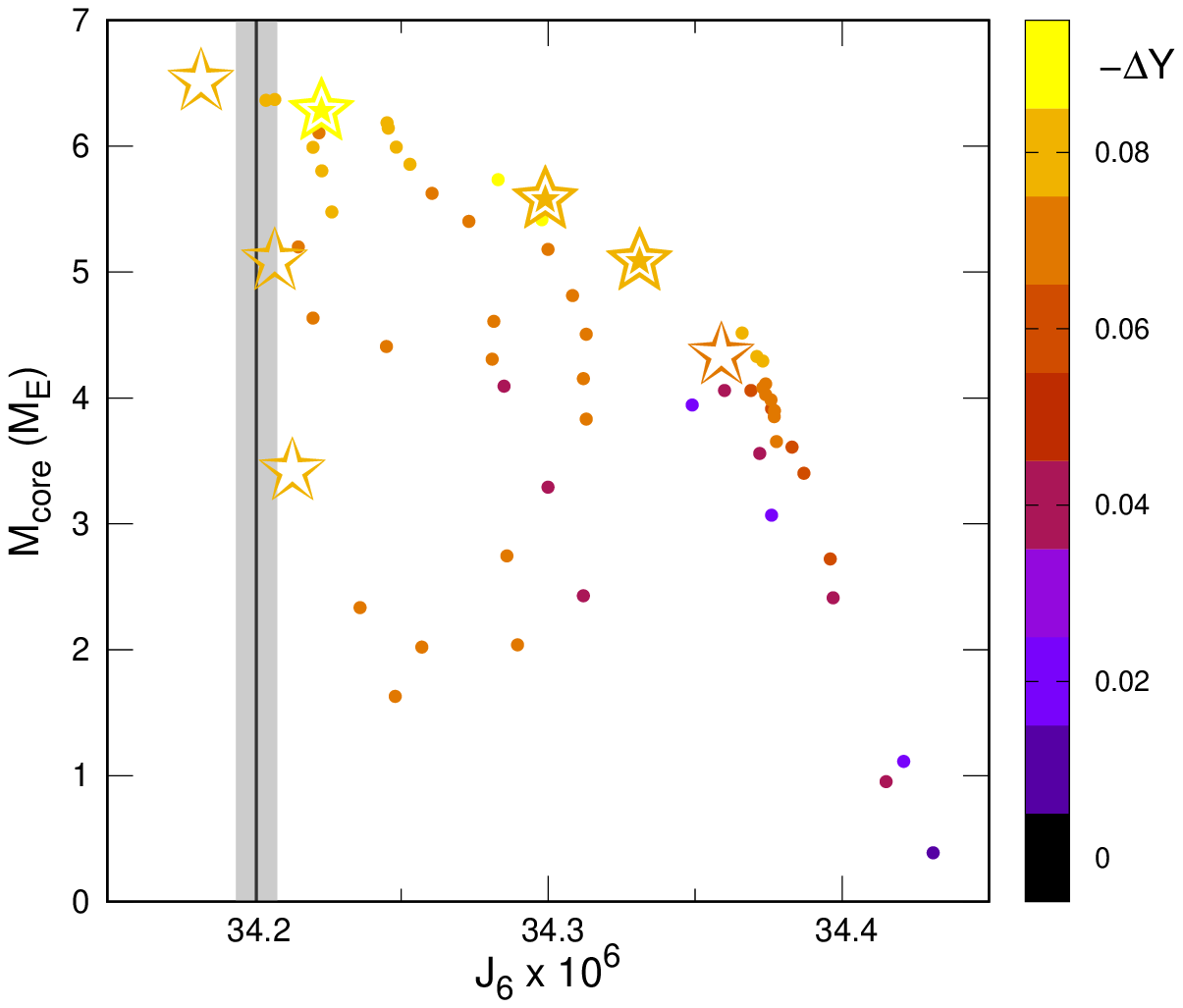}}
\caption{\label{fig:McJ6}
Core mass and $J_6$ values of models with inverted He-gradient (dots) and additional inverted Z-gradient (stars). As in Fig.\ref{fig:ZZ_dYdZ}, open stars show models where $Z_3\lesssim 1\times$ solar while pattern-filled stars show models with $Z_3 \gtrsim 1\times$ solar. The vertical grey bar indicates the Juno measurement of $J_6$. 
The level of additional He depletion $\Delta Y$ in L3 with respect to $Y_{\rm Gal}$ is indicated 
by the color code, with yellow indicating the strongest possible depletion.
}
\end{figure}

Figure \ref{fig:McJ6} shows core mass and $J_6$ values. Solutions with small or zero core mass are a challenge to find for our numerical convergence procedure, but from the trends in the Figure one can guess that more such models exist. With increasing He-depletion, the core mass increases because of the then rapidly decreasing $\Zdeep$ values to fit the $J_s$. At high He-depletion, the model $J_6/10^6$ values approach the observed value 34.201(007) \citep{Durante20}. 

A better fit to $J_6$ implies a smaller wind effect than required for the adiabatic models without an OSL where, when $J_2$, $J_4$, and $\Tone$ and the condition of an about solar  $\Zatm$ are met, $J_6/10^6$ is often found to be 0.2--0.3 higher than observed \citep{Militzer22,Howard23,Militzer24}. Our previous CMS19-EOS \citep{CMS19}-based Jupiter model \citep{N21} that did closely match $J_6$ also yielded a good match to the odd and high-order even harmonics $J_n$. Accordingly, the required deviations from the cloud level wind velocities were with of up to 15 m/s smaller than those required for the ensemble of models of \citep{Militzer22} of 10--50 m/s hat showed a larger deviation in $J_6$.

\begin{figure}
\centering
\rotatebox{270}{\includegraphics[width=0.50\textwidth]{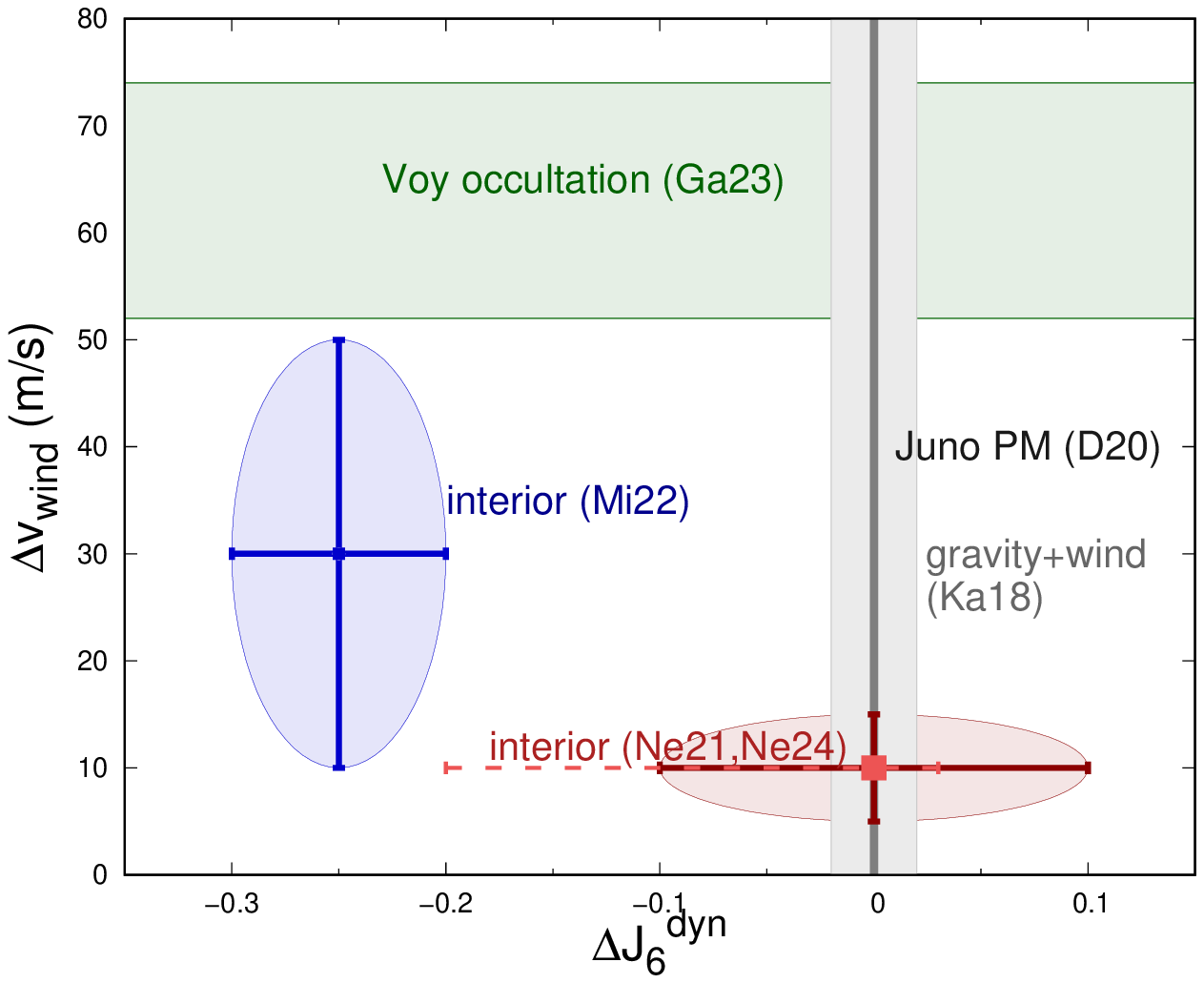}}
\caption{\label{fig:J6winds}
Compilation of absolute wind velocity corrections and dynamical contributions to $J_6\:(\times 10^6)$ that would bring the model values into agreement with the observed values.
Grey vertical bar: observational uncertainty as in Fig.~\ref{fig:McJ6}; red ellipse: $J_6$ model results from this work (Ne24) placed at $\Delta v_{\rm wind}=10\:$m/s and model results from \citet{N21} which led to a deviation from the observed winds by $\lesssim 15$ m/s; blue ellipse:  ensemble of solutions in \citep{Militzer22}; green horizontal bar: wind modifications that are needed to bring the dynamical heights inferred from the zonal winds and Juno gravity data into agreement with the dynamical heights inferred from the Pioneer and Voyager occultation radii \citep{Galanti23}.
}
\end{figure}

However, the wind velocities at the cloud level and at greater depths also have uncertainties. While the wind velocities from cloud tracking obtained with HST have been confirmed by ground-based Doppler spectroscopy \citep{Schmider24}, the agreement is good to within 10-20 m/s. This is of the order of the wind modifications needed by the Jupiter models to fit $J_6$ as described above, although whether those occur at the latitudes as needed is unclear. Larger uncertainties may exist at greater depths, but one most note that they would influence the gravity data, which do not require larger deviations at depth. On the other hand, larger deviations of the order of 60--75 m/s appear to be needed to bring the dynamical heights inferred from the winds into agreement with those inferred from the Pioneer and Voyager occultation data. More accurate or independent occultation radii measurements are desirable to solve the dynamical heights mismatch and to more reliably determine the wind velocities at depth. A compilation of the suggested wind modifications and related dynamical contributions to $J_6$ is shown in Figure \ref{fig:J6winds}.

\subsection{Deep He-depletion and H/He phase separation} \label{sec:res_HHesep}

\begin{figure}
\centering
\rotatebox{270}{\includegraphics[width=0.50\textwidth]{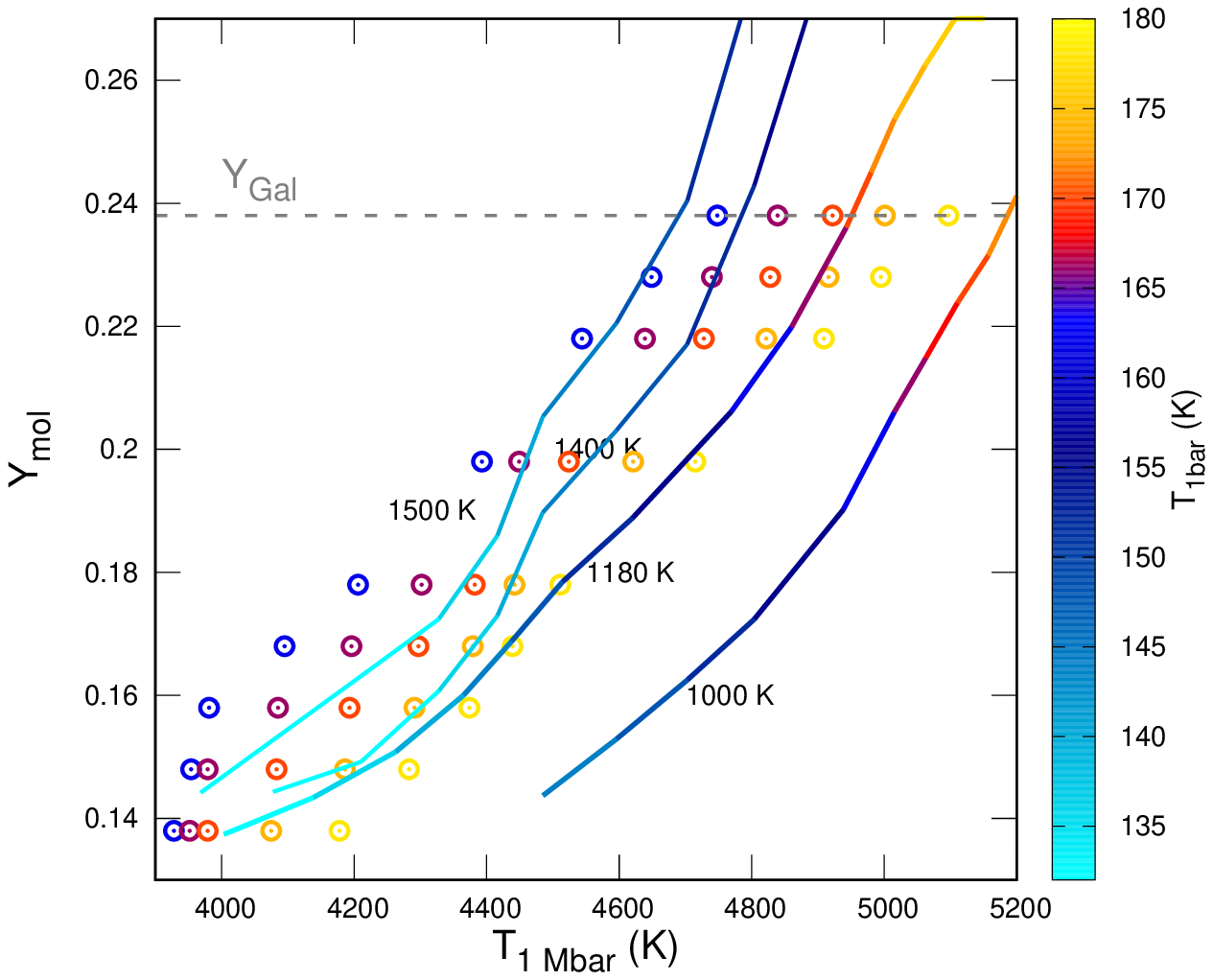}}
\caption{\label{fig:dmxOverlap}
He depletion at the 1 Mbar level. Lines show the  $Y$-$T$ relation at 1 Mbar of adiabatic H/He-profiles according to the LHR0911 H/He-phase diagram for different shifts thereof, while points are for H/He-profiles with sub-adiabatic OSL. The color indicates $\Tone$ of the underlying adiabats (lines) or model P-T  profiles (points). 
}
\end{figure}

The deep He-depletion is assumed to result from H/He phase separation. Here, we compare predictions
from the LHR0911 phase diagram \citep{Lorenzen09, Lorenzen11} shifted in temperature. Consistent solutions occur when the He-depletion level predicted by the (shifted) phase diagram along an adiabat overlaps with the level assumed in the interior models for L3, and where $\Tone$ of the adiabat is less or equal $\Tone$ of the interior model with subadiabatic OSL. We compare the He-depletion level in Figure \ref{fig:dmxOverlap} right at the top of the He-rain zone at 1 Mbar.

For adiabatic standard models ($dY=0$, $\Tone=170\:$K) based on CD21-EOS, a rather fine-tuned shift of this H/He phase diagram by -1180 K is needed to yield the observed atmospheric value $Y_{\rm Gal} \sim 0.238$ \citep{N24}. In models with an OSL, the overlap region is wider and the necessary shift relaxes somewhat to be within -1180 to -1500 K. A larger negative shift means that the phase diagram for a given adiabat leads to less rain-out, while a colder adiabat for a given phase diagram leads to stronger rain-out. A larger negative shift works if the adiabat has low He-abundance and is thus colder, so that both effects balance each other.
 
For a large shift by $\lesssim -1500\:$K, the He-rain region closes even for cold adiabats (cyan), and no overlap is obtained. Conversely, for small shifts $\gtrsim -1000$ K, the then strong He-depletion predicted by the phase diagram cannot be balanced by a warmer planet profile adiabat unless $\Tone \gtrsim 180\:$K (yellow).  
Where overlap is seen, the adiabat can have an up to $-40\:$K lower $\Tone$ than the models with an OSL have. 
This $\Tone$ may be called a `virtual temperature'. Due to the sub-adiabatic stable layer, the observed 
$\Tone$ appears warmer than the virtual temperature of the outward extended  adiabat of the deep interior.
To obtain a cold interior adiabat implies that the heat must have been radiated out more efficiently 
than in case of convection. Hence the OSL must be radiative.

\subsection{Helium-transport through the OSL}\label{sec:diffHe}

In this section investigate under what circumstances it might be possible to have at the top
of the OSL the observed He-abundance $Y_{\rm Gal}=0.238$ while at the bottom a much stronger depletion. 
For this purpose we solve the diffusion equation under simplified, but we believe reasonable 
assumptions, and boundary conditions. 

The diffusion equation of He in a H/He-system can be written
\begin{eqnarray}
\rho\:\frac{\partial Y}{\partial t} 
&=&	\bigg[D_{\rm He} -Y(D_{\rm He}-D_{\rm H}) \bigg]
	\rho\:\frac{\partial^2 Y}{\partial r^2} \nonumber\\
&& + 2\:\bigg[ D_{\rm He} - Y(D_{\rm He}-D_{\rm H}) \bigg]
	\frac{\partial\rho}{\partial r}\:\frac{\partial Y}{\partial r}  \nonumber\\
&& +\: \bigg[ (D_{\rm He} - D_{\rm H}) (Y-Y^2)\bigg]\frac{\partial^2\rho}{\partial r^2} \quad,\label{eq:dYdtDHHe}
\end{eqnarray}
where the third term describes the gravitational settling of He.
According to \citep{French12}, the diffusivities of He and H at 0.98 $\RJ$ are respectively 0.43 and 0.437 mm$^2$/s, very similar. Since our assumed OSL is thin and extends from 0.99 to 0.97 $\RJ$, we assume a constant $D_{\rm He}=0.43\:$mm$^2$/s and set $D_{\rm H}=D_{\rm He}$. This eliminates the gravitational settling term.
Furthermore, we assume constant  $\rho$ for simplicity, which eliminates the second term in brackets,
although $\rho$ changes by an order of magnitude across the OSL. Including the second term may influence
the resulting $D_{\rm He}$ value that is needed to reproduce $Y_{\rm Gal}$ by a factor of a few.
We solve the thus simplified diffusion equation 
\begin{equation}\label{eq:DGL}
\frac{\partial Y}{\partial t} = {\rm D}\: \frac{\partial^2 Y}{\partial r^2}
\end{equation}
using the Crank-Nicolson discretization with $D=f_{\rm enh}\: D_{\rm He}$ with enhancement factors $f_{\rm enh}=$1---1000.
This inhomogeneous 2nd order linear differential equation (\ref{eq:DGL}) can be written $\vec{u}=\hat{A}\vec{b}$, 
where $\vec{u} = u^{j+1}(x_0,\ldots,x_i,\ldots, x_N)$ are the to-be-solved-for $Y$-values at time next step $j+1$
at the locations $x_i$. The sparse matrix $\hat{A}$ contains a diagonal stripe of elements $-\alpha,2(1+\alpha),-\alpha$ 
with $\alpha=D\Delta t/\Delta r^2$, except for $A_{00}=A_{NN}=1$. To obtain $\vec{u}$, we invert 
matrix $\hat{A}$ using the Gauss-Jordan elimination method as described in the Numerical Recipes \citep{NumRep}. 

\begin{figure}
\centering
\rotatebox{270}{\includegraphics[width=0.40\textwidth]{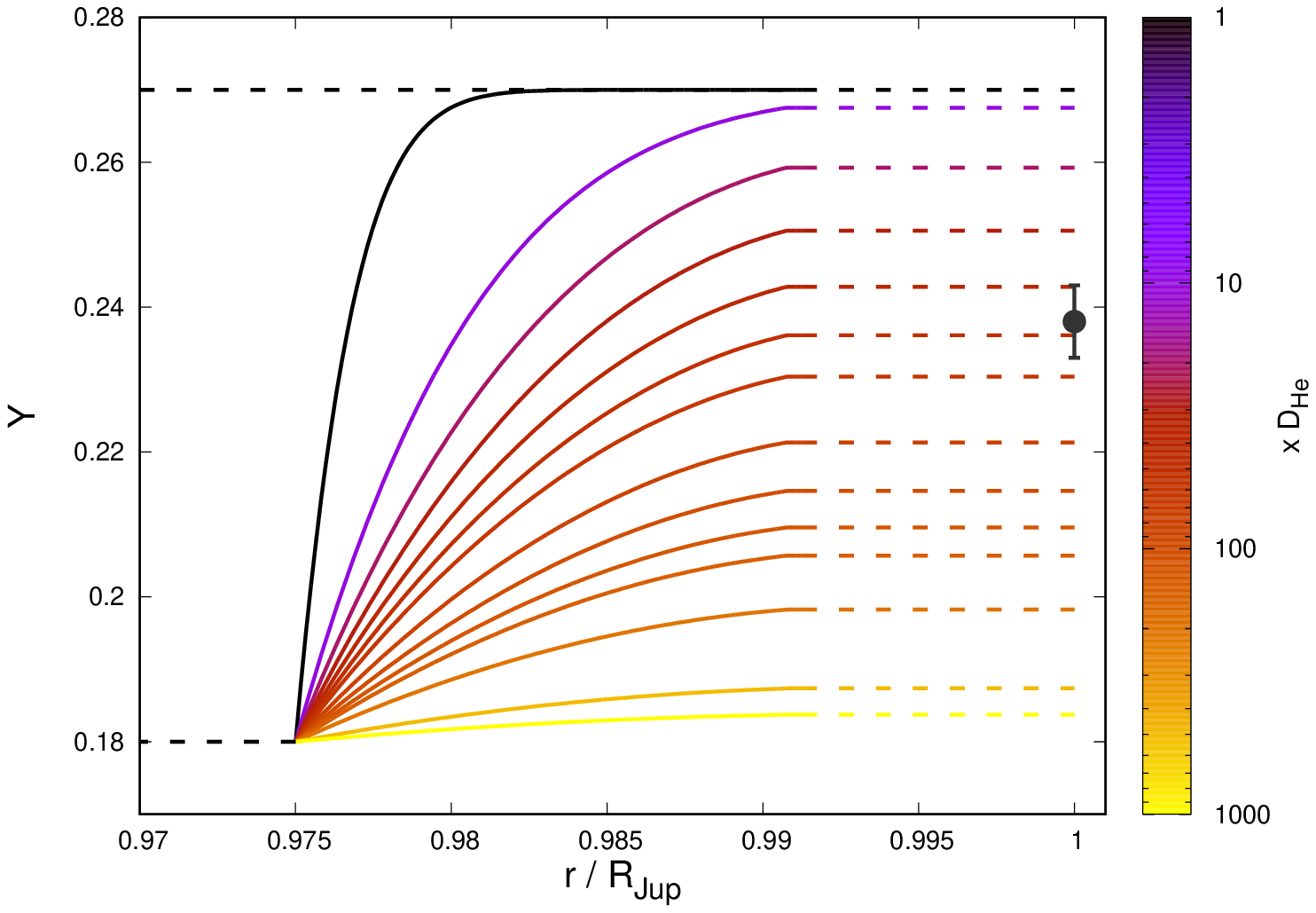}}
\rotatebox{270}{\includegraphics[width=0.40\textwidth]{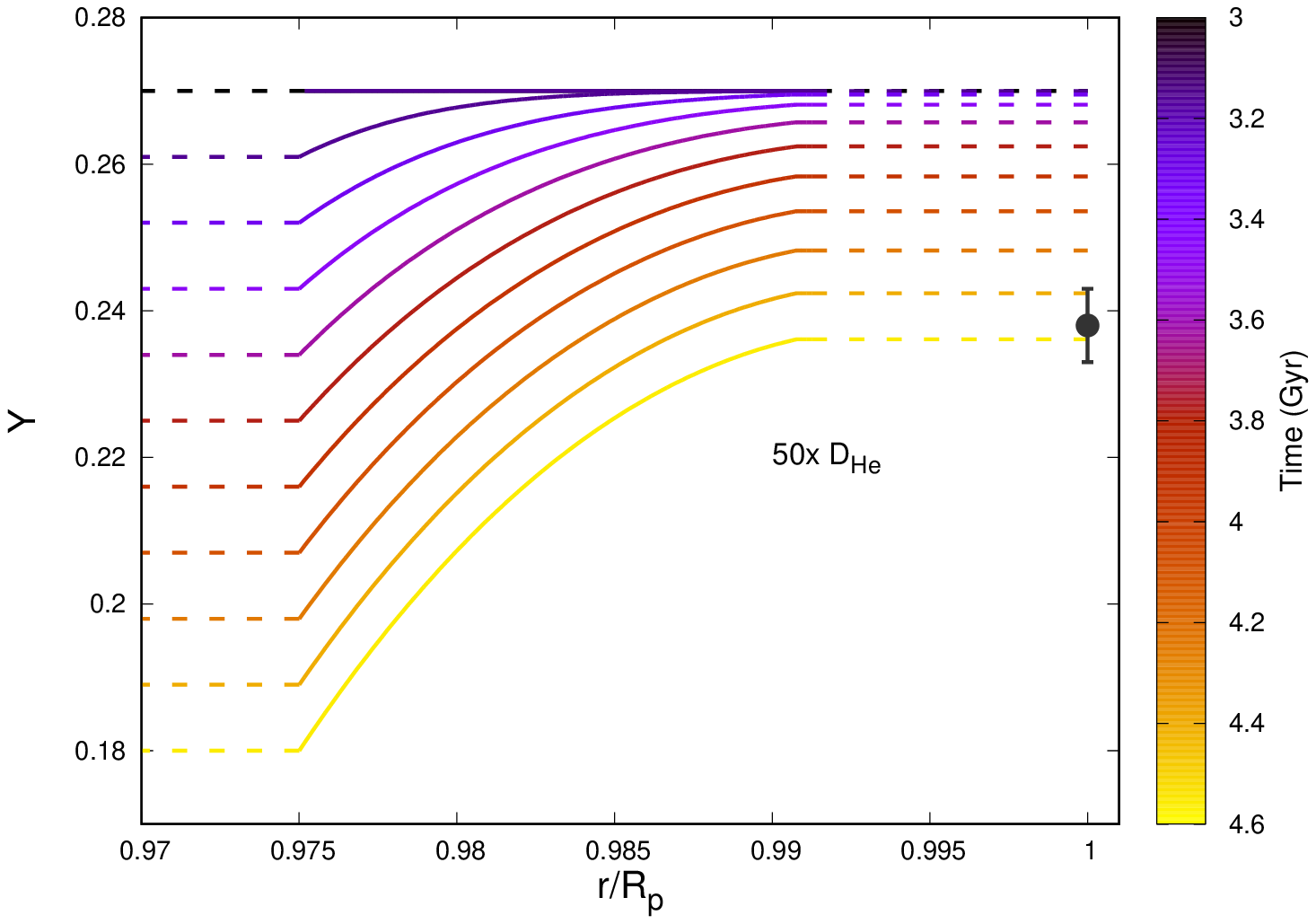}}
\caption{\label{fig:DHe}Diffusion of He through the OSL under simplified assumptions. 
Top panel (a): He-abundance profile at present time for various enhancements $f_{\rm enh}$ (color code) 
of the constant He-diffusivity $\rm D_{He}$; bottom panel (b): He-abundance profiles over time (color code) 
for $\rm D_{He}$ enhanced by a factor of 50 and assuming a linear decrease of $Y(t)$ with time at the bottom 
of the OSL. The black symbol denotes $Y_{\rm Gal}$ at present time.}
\end{figure}

Finally, specifying the elements of the vector $\vec{b}$ requires the two boundary conditions $u_j(x_0)$ 
and $u_j(x_N)$ at the  top and the bottom of the OSL, respectively. At the bottom we impose a monotonous decrease 
of $Y(t)$ over time. We set $Y(t_0)=Y_{\rm proto}$ at $t_0=3.0$ Gyr. For comparison, \citet{N15} obtained an 
onset of demixing in Jupiter depending on the phase diagram modification between 2.5 and 3.7 Gyr, \citet{MF20}
find an onset between 3.5 and 4.0 Gyrs, while \citet{Howard24} between 3.5 and 3.8 Gyr if, in all cases, the phase
diagram is adjusted to $Y_{\rm Gal}$ at the present time. For the monotonic decrease of $Y(x_N)$ in the He-rain zone
and thus at the bottom of the OSL we assume a linear behavior with time with $Y=0.18$ at present, 
but alternatively also consider a quadratic-decrease of Y with time. For the top He-abundance 
we assume that convective overshoot occurs from the convective atmosphere into the OSL. We run our 
diffusion model with overshoot into the first 4 grid points out of 100. This means that we 
equilibrate the computed abundances over these four cells with the atmospheric abundance. Without convective overshoot, 
the atmospheric $Y$ would remain at the initial value at all times. Varying the number of grid points 
between 2 and 8 where overshoot is assumed to occur changes the resulting atmospheric He-abundance by only 1\%.

Figure \ref{fig:DHe}a shows the resulting He-abundance profile for different enhancement factors $f_{\rm enh}$
of the constant He-diffusivity $\rm D_{He}=0.43 mm^2$/s. An enhancement factor of 1 represents the 
purely diffusive case. In that case, the He-abundance in the atmospheres decreases so slowly with time
that even after the assumed 1.5 Gyr of evolution with He-rain, the atmospheric He-abundance would barely
have been changed. The observed atmospheric He-depletion can be explained by an enhancement by a factor
of 50--80. Much higher enhancements by several 100$\times\rm D_{He}$ would deplete the atmosphere in helium rather 
fast. For factors of the order of 1000, the communication between top and bottom of the OSL 
becomes almost instantaneous, as in the convective case.
 
In Figure \ref{fig:DHe}b we fix $f_{\rm enh}$ to a favorite value of 50 and show how the 
He-abundance profile evolves over time. One can clearly see the imposed linear bottom boundary condition
$Y\sim t$. Given the assumptions made, we caution that our estimate of the necessary enhancement factor 
is an order-of-magnitude estimate. Values of several 10 are favored, as opposed to a factor of a few 
or several 100. 

\begin{figure}
\centering
\rotatebox{270}{\includegraphics[width=0.40\textwidth]{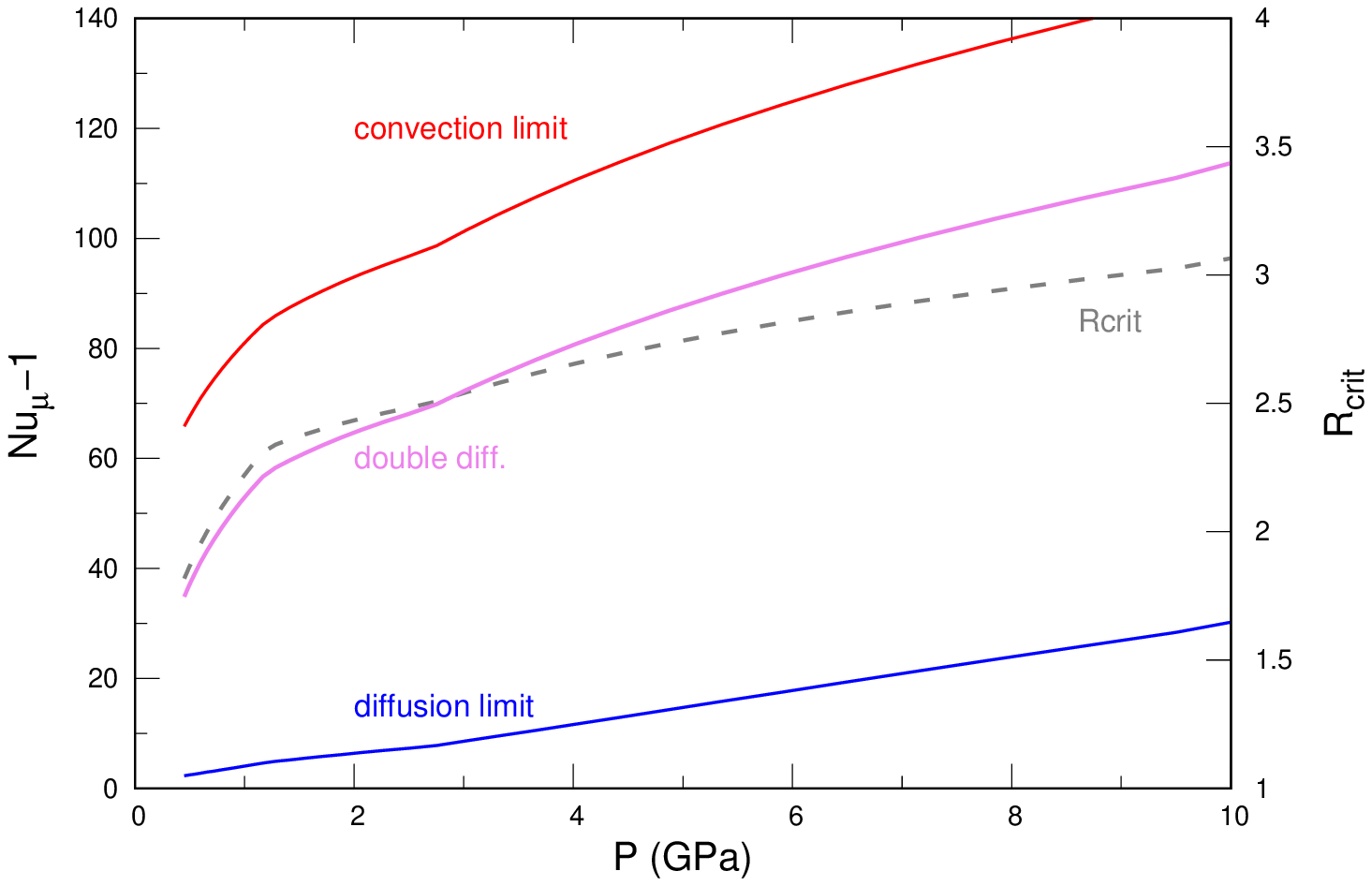}}
\caption{\label{fig:nusselt}
Nusselt number of particle transport in the regime of fingering double diffusive convection according to the
fit formula of \citep{Brown13} for the three values of $R_{\rho}=1$ (red), $R_{\rho}=R_{\rm crit}$ (blue),
and pir choice for the OSL $R_{\rho}=1.11$ (violet). Computation of $Nu_{\mu}$ and $R_{\rm crit}$ involves 
knowledge of thermophysical properties, which depend on pressure (x-axis) and temperature. On the right-hand 
y-axis we plot $R_{\rm crit}$, which apparently adopts values between 1.5 and 2.5 across the OSL (0.1--2 GPa).}
\end{figure}

Particle transport enhancement factors of the order 50 place the OSL in the regime of fingering 
double diffusive convection, which is consistent with the constraint on the density ratio $R_{\rho}^{-1}$=0.9 
we made to derive the sub-adiabaticity and favored bottom He-abundance today.  
\citet{Brown13} obtain particle transport enhancements of the order of 5--90 for the fingering double diffusive
regime, which they parameterize by a Nusselt number $Nu_{\mu}-1$. We display their fit formula for $Nu_{\mu}(R_{\rho})-1$ 
as a function of the density ratio $R_{\rho}$ in Figure \ref{fig:nusselt} for the three values
$R_{\rho}=R_{\rm crit}$, which marks the transition to the diffusive regime, $R_{\rho}=1$, which marks the transition
to overturning convection, and our choice $R_{\rho}=1.11$. According to Figure \ref{fig:nusselt}. the latter choice 
yields $f_{\rm enh} =$30--60 across the OSL. A less conservative choice $R_{\rho} \sim 1.05$ 
would imply a faster particle transport. 
Given current uncertainties, the prediction of the efficiency of particle transport in fingering double diffusive
convection and what is needed to transport He through the OSL are consistent.
To compute $Nu_{\mu}$ and $R_{\rm crit}$,  we applied the material properties along the Jupiter adiabat of \citet{French12}, 
which begin at P = 0.45 GPa.
Clearly, a better understanding of particle transport in the $\sim$0.1 GPa (1 kbar) region is needed
in order to place tighter constraints on the possibility on a double-diffusive OSL.

\subsection{Neon transport through the OSL}

It has been suggested that neon prefers He-rich droplets over a metallic-hydrogen environment 
to mix with \citep{WiMi10} and would therefore appear depleted in Jupiter's atmosphere if H/He 
phase separation occurs. Indeed, the Galileo entry probe measured a strong depletion in neon 
of Ne/H$_{\rm Gal}=0.123\times 10^{-4}\pm 10\%$ \citep{Atreya03}, to be compared to a protosolar value 
Ne/H$_{\rm proto}=1.23\times 10^{-4}$ \citep{AndGrev89}, or $1.73\times 10^{-4} \pm 26\%$ \citep{Lodders21}. 
With these reference values, Ne/H in Jupiter's atmosphere is only 0.10 times (0.07 times) the 
protosolar value. For He/H in Jupiter's atmosphere we derive from $Y_{\rm Gal}=0.238$ a He/H$_{\rm Gal}=0.0781$, 
yielding Ne/He$_{\rm Gal}=1.575\times 10^{-4}$. With a protosolar He/H$_{\rm proto}=0.099$ \citep{Lodders21} 
($Y_{\rm proto}=0.2836$), Ne/He$_{\rm Gal}$ is only 0.090 times the protosolar Ne/He. 
Conversely for $Y_{\rm proto}=0.270$, He/H$_{\rm proto}=0.09245$, and using furthermore  the
lower Ne/H$_{\rm proto}$ of \citet{AndGrev89}, Ne/He$_{\rm Gal}$ would be 0.118 times the protosolar Ne/He. 
The uncertainty is 41\% due to Ne/H$_{\rm proto}$ (26\%), He/H$_{\rm Gal}$ (5\%), and Ne/H$_{\rm Gal}$ 
(10\%). 

As for helium, we assume that the loss of Ne at the bottom of the OSL is the same as that in the He-rain region. 
\citet{WiMi10} provide a relation between the loss rate of neon and the loss rate of helium,
\begin{equation}
\frac{dX_{\rm Ne}}{dt} = X_{\rm Ne}\exp\left(\frac{\Delta G_{tr}}{k_BT}\right) \frac{dX_{\rm He}}{dt}
\end{equation}
where $X_i$ is particle concentration and $\Delta G_{tr}$ is the Gibbs free energy difference between
placing a Ne particle into the metallic hydrogen while keeping the He-droplets free of neon, 
or transferring Ne into the helium droplet while keeping the H-environment free of Ne.
For $P$--$T$ conditions where He-rain is expected to occur, \citet{WiMi10} obtain a range 
$-3.24 \leq \Delta G_{tr} \leq -1.14$ eV, with the $-$sign indicating preference for partitioning
into helium droplets. Here we use values $-3.4 \leq \Delta G_{tr} \leq -1.6$ eV. \citet{WiMi10} find
that the neon partitioning preference into He droplets decreases with temperature, and preferences weaker 
than $-1.90$ eV would require low temperatures less than 4000 K in the demixing region.

\begin{figure*}
\begin{minipage}{\textwidth}
\includegraphics[angle=270,width=0.60\textwidth]{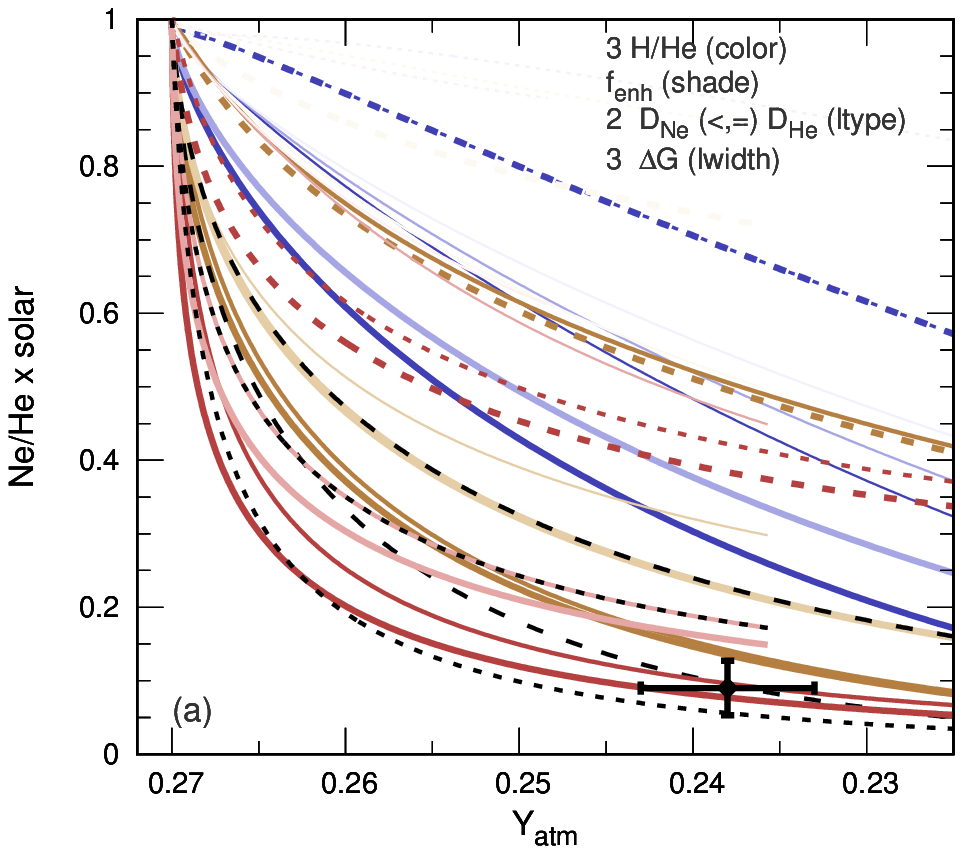}
\hspace*{-3cm}
\includegraphics[angle=270,width=0.60\textwidth,clip=true]{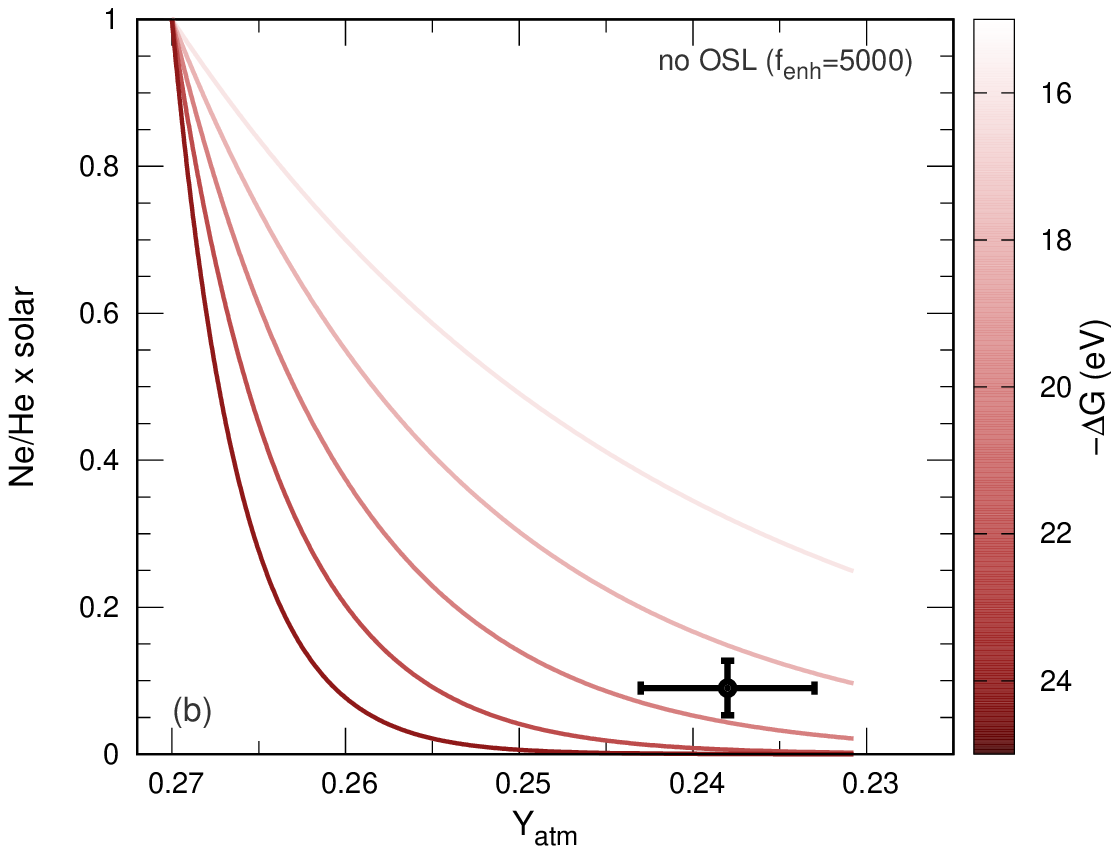}
\end{minipage}
\caption{\label{fig:nehe}Evolution of the atmospheric Ne/He ratio and the atmospheric He mass abundance 
due to the ongoing He-rain underneath the OSL (a) or in the absence of an OSL (b)
At time $t_0$ when He rain begin, all curves begin at Ne/He=1.
Left panel: Different colors represent different H/He phase diagrams via the 
exponent $n=0.5$ (blue), 1 (yellow), 2 (red).  Color intensity increases from $f_{enh}=50$ 
to 140. Solid lines are for $D_{\rm Ne}=D_{\rm He}$ while colored dashed lines for 
$D_{\rm Ne} < D_{\rm He}$. Thin, medium, thick lines are for $-\Delta G=-1.6, -2.8, -3.4$. 
Free black dashed lines assume $f_{enh}=80$ and $-\Delta G=2.8$ but a thinner OSL 
compared to the respective solutions indicated by black dashed ontop colored lines of same 
line-style. Right panel: Without OSL, therefore  curves for different exponents $n$ 
fall on-top of each other. Color intensity indicates the Ne partitioning preference $\Delta G$. 
The black symbol indicates the observed value in Jupiter's atmosphere.}
\end{figure*}

In addition to $\Delta G_{tr}$, we explore four further parameters. One is the Ne diffusion coefficient.
\citet{Wilson15} finds that the diffusion coefficient of a species in a H-He environment at Mbar pressures 
scales with the atomic mass $m_i$ as $ D_i\sim D_H\:\sqrt{m_H/m_i}$. However, at the much lower 
pressures and temperatures around the OSL, \citet{French12} find $D_{\rm He}\sim D_{\rm H}$. 
We assume either $D_{\rm Ne} = D_{\rm He} \sqrt{m_{\rm He}/m_{\rm Ne}}$ (low), or
$D_{\rm Ne} = D_{\rm He}$ (high) (solid/dashed lines in Fig.\ref{fig:nehe}a)). The latter case may also be a result 
of macroscopic mass transport in the double diffusive regime compared to single particle diffusion in the diffusive case. Another parameter is the He diffusivity enhancement factor $f_{enh}$, which we vary from 50 to 140 (shade in Fig.~\ref{fig:nehe}a).

We also vary the imposed relation for the He-depletion rate $dX_{\rm He}/dt$ at the bottom 
of the OSL for which we assume a power law in the form $Y_{\rm bot}(x)=a_0 + a_1\:x^n$ with exponent $n=0.5$, 1, 2 (colors in Fig.\ref{fig:nehe}a) and $a_0, a_1$ adjusted to $Y_{\rm bot}(0)=0.27$, $Y_{\rm bot}(1)=0.18$,  and $x(t)$ mapping time linearly from $t_0=3\:$Gyr, $\tau=4.56\:$Gyr to [0,1]. Variation of the exponent $n$ is supposed to represent different H/He phase diagrams that would yield the assumed relations for $Y_{\rm bot}(t)$. The fifth's parameter we vary is the thickness of the OSL, which we reduce as an example from 0.1-2 GPa to 0.1--1 GPa (black dashed lines in Fig.\ref{fig:nehe}a).

Apparently, it is challenging, but not impossible, to explain the observed low Ne/He$_{\rm Gal}$ of 0.09 times the protosolar value in Jupiter's atmosphere in the presence of an OSL. The observed Ne/He can only be met if Ne was washed out quickly with the beginning of He-rain (low $\Delta G_{tr} \lesssim -2.8$, medium-to-thick lines), if Ne is transported very efficiently (solid lines) with the diffusivity near the upper limit of the  predictions from double diffusive convection ($f_{\rm enh}\gtrsim 80$) (intense colors), and if He-rain proceeded slow in the past $n\geq 1$ (orange, red) and not that strongly ($Y_{\rm bot} > 0.18$ as otherwise, with the high $f_{\rm enh}$ values, $Y_{\rm atm}$ would today be lower than observed.)
Notably, strict agreement within the externally provided constraints ($f_{enh} \lesssim 80$ \citep{Brown13}, $\Delta G > -3.1$ \citep{WiMi10}) is obtained (free black dashed lines in Fig.~\ref{fig:nehe}a) if the OSL is slightly thinner (e.g., 0.1--1 GPa/0.99--0.98 $\RJ$) than for our default values of 0.1--2 GPa/0.99--0.975 $\RJ$. 
Taken at face values, the low observed Ne/He would be indicative of a thin OSL 
$\Delta r\lesssim 0.01 \RJ$ that ends above 10 kbar (2100 K). 

That the low observed Ne/He suggests that an OSL should be thin poses the follow-up question 
if this suggests the absence of an OSL? Figure \ref{fig:nehe}b shows the evolution of Ne/He in the 
absence of an OSL, here numerically represented by convection-like high diffusion coefficient 
($f_{enh}$)5000). In this case, the atmosphere becomes quickly depleted in Ne unless the Ne partitioning 
preference into He-droplets is assumed to be moderate. A fine-tuning  $-2.0 < \Delta G < -1.8$ is required.  
This is marginally consistent with the finding $-2.8 < \Delta G < -1.9$ by \citet{WiMi10}, where their 
upper limit is based on an otherwise unexplained cold He-rain region of less 4000 K. The thin overlap region 
implies a required fine-tuning in the Ne-partitioning preference of $-2.0 < \Delta G < -1.9$ while 
independence on the assumed H/He phase diagram as the depletion at depth is quickly reported
to the atmosphere. We see this behavior numerically when setting $f_{enh}$ to a convection-like 
high value of 5000.  We conclude that in both cases, with or without an OSL, the observed Ne/He can be 
explained but it requires some fine-tuning. Importantly, the required Ne partitioning preferences 
differ substantially, with $-\Delta G= 1.9$ to 2.0 in the absence of an OSL while $\gtrsim 2.8$ 
in the presence of an OSL. An independent determination of this parameter from theory or experiment 
would therefore be highly valuable for discriminating between models without and with a moderately 
thick OSL.

\section{Discussion}\label{sec:discuss}

\subsection{Opacity in the OSL} \label{sec:dis_opac}

We assumed that the physical reason for the existence of the OSL is that the heat can be transported
out by radiation. Here, we compare the sub-adiabatic temperature gradient in the OSL that we constrained
by assuming a double-diffusive stability criterion $R_{\rho}^{-1}=0.9$ to the radiative gradient
\begin{equation}
\nabla_{\rm rad} = \frac{3}{16\pi\,acG}\:\frac{\kappa_R\:L\:P}{m\,T^4}
\end{equation}
where $a= 7.57\:10^{-16}\:$J/m$^3$/K$^4$ is the radiation constant, $c$ the speed of light, 
$\kappa_R$ the Rosseland mean opacity, for which we use the tabulated values of \citep{Freedman08} 
for solar composition with extrapolation toward higher pressures, $L$ is the intrinsic luminosity 
from the interior, and mass $m$, pressure $P$, temperature $P$ are taken from the planetary profile. 
The intrinsic luminosity $L\sim T_{\rm int}^4$ is determined from the observed effective temperature 
$T_{\rm eff}$ and the Bond albedo $A_B$ according to $T_{\rm int}^4 = T_{\rm eff}^4 - (1-A_B)/4\: (R_{\odot}/a_{\rm orb})^2$. 
The two different Albedo values from Voyager, $A_B=0.34$ and Cassini, $A_B=0.50$ \citep{LimingLi18}, 
yield two different intrinsic temperatures. 

\begin{figure}
\centering
\rotatebox{270}{\includegraphics[width=0.5\textwidth]{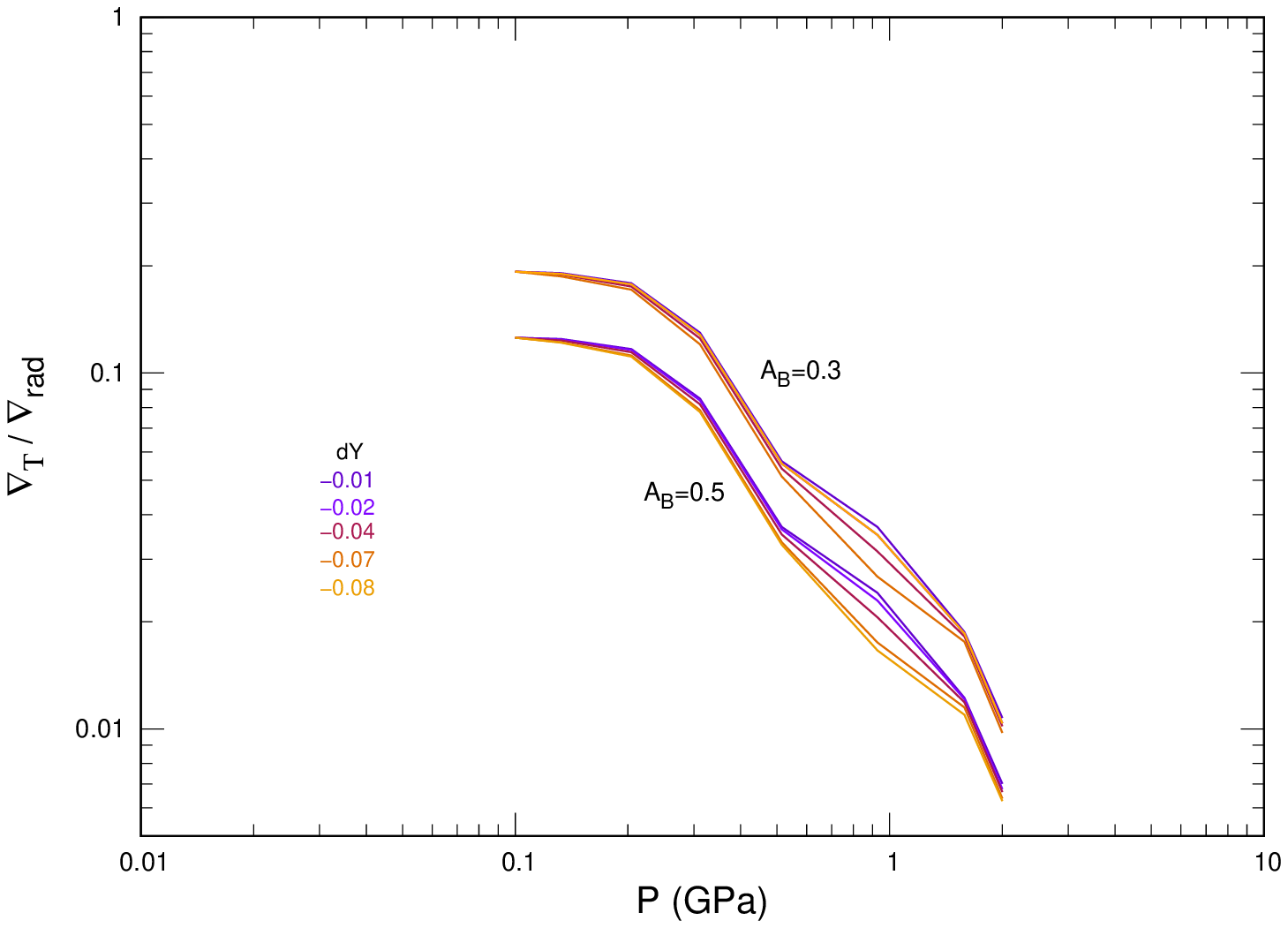}}
\caption{\label{fig:grads_kR}
Ratio of thermal to radiative gradient in the OSL. This ratio directly yields the required reduction in Rosseland mean opacity for solar composition $\kappa_R\sim \nabla_{\rm rad}$ for the OSL to be radiative. Curves are for different assumed $\Delta Y$ values (color code) and the two different Bond albedo measurements. }
\end{figure}

For the OSL to be radiative, the ratio of thermal gradient $\nabla_T$ to radiative gradient 
$\nabla_{\rm rad}$ must be larger than 1. Figure \ref{fig:grads_kR} shows this ratio for
$\nabla_{\rm rad}$ computed with the Rosseland mean opacity for solar composition. 
The ratio is 0.01--0.2. Since $\nabla_{\rm rad}\sim \kappa$, this ratio equals the reduction in 
Rosseland mean opacity for the OSL to be radiative. The needed reduction in opacity is a factor
0.01--0.2 of that for solar abundances. 

\citet{Guillot94} found that a radiative windows opens between 1300 and 2700 K along the Jupiter 
adiabat if metals and volatiles are permitted while alkali metals are not, leading to a reduction
in Rosseland mean opacity with respect to the critical value where convection would set in 
($\nabla_{\rm rad} =\nabla_{\rm ad}$) to a fraction $\gtrsim 0.03$. Since our estimate is of same
magnitude, this comparison suggests that absence of alkali metals might be required to reduce
the opacity sufficiently.

From the limb darkening and brightness temperature observed by Juno's MWR instrument in the longest 
wavelengths channel corresponding to 0.6 GHz, which is sensitive to the emission at pressures of 
about 100--200 bars,  \citet{Bhatta23} concluded an alkali metal depletion to a fraction
$10^{-2}$--$10^{-5}$ with respect to solar abundances. The electrons from the remaining abundance 
of the ionized alkali metals would be sufficient to explain the observed excess opacity over that 
expected if only water and ammonia were present.  They suggest that alkali depletion to that extent 
may be indicative of a radiative zone and that, if the excess opacity is due to other species, 
the alkali metal depletion down to 1 kbar could be even stronger. However, one must caution that
the observationally inferred opacity refers to 100-200 bars while the opacity at the level of 
1000--10,000 bars where alkali metal depletion can cause a radiative window remains
observationally unconstrained.

\citet{Muller24} show that the low opacity that leads to the radiative window found by \citep{Guillot94} 
can be represented by a reduction of $\kappa_R$ to a fraction of $\sim$ 0.1. They confirm 
the appearance of a radiative window for present Jupiter at around 2000 K, 1-10 kbar 
and find that it could persist over geological timescales while it moves inward when the planet cools.  
While these studies support the assumption of a radiative zone at the pressures where the interior
models of this work favor it, we caution there is still no firm evidence of its existence.

In passing, we note that that evidence is mounting for an accelerated cooling of Jupiter. 
A subadiabatic OSL, like a radiative window \citep{Guillot94}, implies that the 
interior undermeath is colder, which physically results from the more efficient heat transport
in the stable layer as otherwise, it would become convective. The colder interior today implies
that a larger fraction of the primordial heat must have already escaped the interior, 
thus the cooling is accelerated. Another evidence comes from 
an enhanced albedo according to Cassini measurements. As the two albedo estimates
from the Voyager flyby and the Cassini flyby disagree and yield different cooling times for Jupiter
\citep{MF20}, a re-assessment of the Jupiter's Bond albedo would be helpful, 
perhaps from Juno measurements.

\subsection{Other evidence for an outer stable layer: zonal winds}

Alternative evidence for a stably stratified layer (SSL) far out in Jupiter's envelope comes
from the zonal winds. 
The Juno Prime Mission goal to infer the depth of the zonal winds from gravity measurements
has been achieved, with the result of a wind-depth of 2000-3000 km possibly depending on 
latitude \citep{Kaspi18,GalantiKaspi21}. The zonal flows are furthermore found to extend along
cylinders \citep{Kaspi23} and to decay radially \citep{Galanti21}.

Hydrodynamic simulations of flows in an outer convective shell reproduce the strong prograde 
equatorial zonal jet. The depth where it decays marks the tangent cylinder (TC). The TC separates 
the flows in the two hemispheres. Within the TC, north-south asymmetric flows can occur at mid- to high latitudes. 
However, hydrodynamic simulations of rotating convective shells fail to produce high-latitude zonal flows, 
see \citet{Wulff22} for an overview. An assumed stably stratified spherical layer (SSL) with a top radius defined by 
the tangent cylinder finally yields high-latitude flows within the tangent cylinder \citep{Wulff22}.
When magnetic effects are neglected, their decay profile is softer \citep{Wulff22} than 
the radial decay profile that is inferred from the Juno gravity data \citep{Galanti21}. 

At the depth where the zonal flows decay, i.e.~beyond a depth of 2000 km, the electric conductivity rises so that
magnetic effects are possible. Consequently, current explanations why the winds decay there involve 
magnetic effects in a weakly conducting environment that is stably stratified \citep{Christensen20,Wulff24JGR}. 

A zonal flow decay depth of 2000--3000 km corresponds to 0.97--0.96 $\RJ$, thus slightly deeper than 
our OSL (0.99-0.975 $\RJ$). According to our interior models,  pressures at 0.97--0.96 $\RJ$ are in the 2.5--8 GPa range. 
Figure \ref{fig:optimOSL} suggests that a stable layer at these pressures would act neutrally on the density at depth. 
Therefore, the zonal flows may exist throughout our non-conducting OSL and just lead to a slightly softer onset
of the decay with depth, perhaps in between the decay profiles constrained by gravity and those constrained
by gravity and the magnetic field \citep{GalantiKaspi21,N21}.  Clearly, more work needs to be done to understand 
the zonal flow in a electrically non-conducting, thin stable layer between the outer convective zone where they are 
generated and the electrically conducting region, where they sharply decay.

\subsection{Other evidence for an outer stable layer: tidal response}

Further alternative evidence of an outer stably stratified region in Jupiter is from its observed tidal
response to Io. Analysis of Juno observations \citep{Durante20} revealed that Jupiter's Love number $k_{42}$ differs
from the hydrostatic value by at least -15\% within its $3\sigma$ observational uncertainty. 

\citet{Idini22} explain this tidal response of Jupiter by the presence of a stable layer that is 
at least $\Delta R = 0.1\RJ$ thick and extends out to $R_o=$0.7--0.8 $\RJ$. 
In this stable layer, g-modes of the tidal frequency of Io can exist and be excited 
by the orbiting satellite. If the outer boundary $R_o$ were deeper/farther out, only higher/lower 
mode frequencies would be permitted that are away from the resonant frequency of Io and therefore
will not be excited. This behavior can be understood by the approximation 
\citep{Idini22} ${}_l^m\omega_g\sim \:\sqrt{l(l+1)}/\pi\:n\:\int_{R_i}^{R_o} dr (N/r)$ 
for a large number $n$ of nodes in the radial component of the mode eigenfunctions. Here, $N$ is the
Brunt-V{\"a}is{\"a}l{\"a} frequency in the stable layer, $R_i$ is the inner boundary radius, and $l$ and $m$ are
degree and azimuthal order of the mode. For constant $N$, ${}_l^m\omega_g\sim N \:\Delta R/R_o$. 

While many modes can contribute to the Love number $k_{42}$, out of the g-modes, only the ${}_4^2g$ mode 
falls into the right frequency range of $\omega_{\alpha}/\Omega_{rot}\sim -1.5$ around the tidal frequency 
of Io and does not simultaneously perturb $k_{22}$ too much \citep{Idini22}. 
In order to excite the ${}_4^2$g-mode over geological timescales rather than in passing, the orbital 
frequency of Io must evolve at the same rate as this g-mode. The requirements for such resonant locking 
seems to be fulfilled for the Jupiter-Io system. 

To summarize, the scenario proposed by \citet{Idini22} imposes the constraint on Jupiter's interior that a stable layer 
of thickness $\Delta R \gtrsim 0.1\RJ$ exists with an outer radius of 0.7--0.8 $\RJ$. Pressures there 
are 2--5 Mbars. Such a location is not supported by the $Z(m)$ profiles of our CD21-based Jupiter models 
with inverted He-gradient. It can not be excluded, though, that H/He phase separation could extend 
over this pressure range. In a strongly super-adiabatic He-rain region however, the He-gradient zone 
would become narrow. Recent Jupiter models with He-rain suggest the He-gradient zone to be at 2-3 Mbars 
\citep{Howard24} and thus be too narrow (0.81--0.77 $\RJ$). As the $Z$-gradient in Jupiter models 
with MH13-EOS or REOS.3 is at the location predicted by \citet{Idini22}, solving this tension may imply that
the true H/He-EOS differs from CD21-EOS.
 
However, the mode frequencies depend on the rotation rate of the planet. \citet{Lai21} and \citet{YLin23} 
point out that the perturbative treatment of rotation used by \citet{Idini22} is insufficient. 
In a rotating planet, inertial modes are raised in response to the Coriolis force. 
Inertial modes can couple to g-modes of similar frequency and shift or broaden the frequency 
range where resonances occur. To account for inertial modes and gravito-inertial mixed modes requires 
a non-perturbative treatment of the rotation-induced Coriolis term. \citet{Lai21} employed a spectral code 
that solves the equations of hydrodynamics directly and found that inertial mode contributions to $k_{22}$ 
are negligible, but their potential contribution to $k_{42}$ were not addressed.

\cite{YLin23} addressed the influence of inertial and gravito-inertial modes on $k_{42}$ by solving
the linearized equations of hydrodynamics including Coriolis force and viscosity but he neglected the centrifugal force, 
which causes rotational flattening. \cite{YLin23} found that under the assumption of an extended dilute core, 
a gravito-inertial wave of frequency $\omega_{\alpha}$ occurs that could influence $k_{42}$ by the desired order 
of magnitude of $\geq 10$\% but that this wave would occur too far away, at $-\omega_{\alpha}/\Omega_{\rm rot} \sim 1.2$, 
from the tidal frequency of Io to be excited. On the other hand, the N values adopted by \citet{YLin23} are up 
to 50\% lower than in \citet{Idini22}. Because of $\omega_{g}\sim N$,  this difference may account for part of the 
difference in the predicted mode frequencies between \citet{Idini22} and \citep{YLin23}. Nevertheless, the work by
\citet{YLin23} shows that a proper treatment of rotation is important to infer reliable conclusions on the presence
of a stable layer from the tidal response.

Finally, \citet{Dewberry23} offers a comprehensive treatment of rotation. The equations of fluid dynamics 
are solved in non-spherical coordinates and $N$ values as in \citet{Idini22} are adopted for the dilute core. 
\citet{Dewberry23} recovers a resonance around Io's tidal frequency as discovered by \citet{Idini22}.
The resonance is found to be broad due to inertial-/g-mode mixing as discovered by \citet{YLin23}. 
Moreover, \citet{Dewberry23} suggests that ${}^2_4$g-modes at different stable layer depths may be responsible 
for the observed dynamical contribution to $k_{42}$ because their amplitude could be enhanced due to excitation 
by the stronger quadrupole-component $U_{22}$ of the tidal potential, rather than only by the weaker $U_{42}$-component.
This possibility offers a solution to reconcile the CD21-EOS based Jupiter models with the observed $k_{42}$ value.

Clearly, more work needs to be done to understand the origin of the observed tidal response of Jupiter that shows
up in the $k_{42}$ value  and the relation to stable layers in Jupiter.

\section{Summary}\label{sec:summary}

The vast majority of Jupiter models that has been constructed to fit the tight Juno gravity data 
fails in at least one of the properties of (i) consistency with observed 1-bar temperature, 
(ii) a minimum of 1$\times$ solar atmospheric Z, or (iii) be derived from a H/He-EOS without artificial
perturbation thereof. One may circumvent this tension problem by arguing for a sub-solar $Z$ in Jupiter's
atmosphere and outer envelope, or for $\Tone$ beyond the $3\sigma$ observational value, or for 
a large (10\%) overestimation of density in current DFT-MD-based H/He-EOS in the 10-50 GPa region. 
Substantially higher $\Zatm$ value of the order of 2$\times$ solar or more would require even stronger 
perturbations or hotter adiabats. 

In this work we investigated how much  Jupiter models can gain in $\Zatm$ under the assumption of an 
Outer Stable Layer (OSL) with an inverted He-gradient and optional inverted Z-gradient.  All models employ the 
CD21 H/He-EOS and assume $\Tone=170\:$K consistent with the observed near-equatorial 1-bar temperatures.
In brief, we find that 
\begin{enumerate}
\item
Favorable locations for the OSL occur where the thermal expansion coefficient $\alpha_T$ is high,
which is the case at around 0.5-10 kbars. We used 1-20 kbars here.
\item
The inverted He-gradient across the OSL leads to atmospheric heavy element abundances 
that are up to $\Delta \Zatm =0.03$ $(+2\times)$ solar higher than for adiabatic models. 
With an additional inverted Z-gradient, $\Zatm$ up to $3\times$solar is possible. 
Atmospheric water abundances O/H$>4\times$ solar remain out of reach.
\item
Models with 1$\times$ solar $\Zatm$ have a dilute core confined to the inner 0.2--0.3$\MJ$
(0.4--0.5$\RJ$) and are consistent with both delayed and non-delayed phase-2 core accretion formation models
\citep{Helled22,Helled23}.
\item
Models with 2$\times$ solar $\Zatm$ suggest a homogeneous-Z interior and that 
the He-rain region is super-adiabatic, in order to prevent $\Zdeep$ from becoming too low.
\item
Models with 3$\times$ solar $\Zatm$ have a 1$\times$ solar homogeneous-$Z$ interior underneath the OSL and a small compact core.
These model are consistent with the classical core accretion formation. Otherwise, they would require
accretion of sub-solar-$Z$ gas and completed erosion of an initial extended dilute core.
\item
The OSL has been in a state of fingering double diffusive convection over the entire time of H/He phase separation.
Heat is transported outward by radiation at a mean opacity that is 0.01-0.2x lower than the Rosseland
mean for solar abundance.
\item
The low observed atmospheric Ne/He ratio suggests that Ne is transported through the OSL 
as efficiently as He is, with an enhancement of $\sim$80$\times$ over the He particle diffusivity,
and that the OSL is thin ($\sim 0.01 \RJ$) and Ne partinioning into He-droplets is efficient. 
In contrast, the absence of an OSL would require distinctively weaker Ne partitioning as 
otherwise, the atmosphere would appear even more depleted.
\item
Our model furthermore relies on a strong He-depletion $Y\sim $0.16 due to H/He-phase separation. 
We showed that this is not an obstacle, because a phase diagram that explains the Galileo value
would lead to correspondingly stronger depletion along a colder adiabat. 
\item
Our models do not support a stable layer below 0.7--0.8 $\RJ$ as has been suggested in order to explain the
dynamically enhanced tidal response of Jupiter's $k_{42}$ value to Io. 
\item
An OSL far out in the non-conducting region and an adjacent stable layer (SSL) in the weakly electrically conducting
may interfere neutrally. The sub-adiabatic OSL may act to initiate a soft slow-down of the winds, while the SSL 
is still sufficiently far out to act neutrally on the deep density profile where the gravitational harmonics
$J_2$ and $J_4$ are sensitive.
\item
To better understand Jupiter's interior requires to know the global temperature profile, the global water abundance,
the H/He phase diagram and the H/He-EOS at pressures of 10-100 GPa, the opacity in the 1-10 kbar region, 
Jupiter's Bond albedo, and the frequencies, amplitudes of normal modes and inertial modes in a rotating fluid planet
with stable layers.
\end{enumerate}

While we caution that in general, there is no one-to-one correspondence between a wanted parameter and the structure and formation of Jupiter, we try to outline how the wanted parameters relate to the desired information:
The H/He phase diagram, the latent heat released upon He droplet condensation, and the thermal conductivity in the He-rain region influence its temperature profile, thickness, and the location of a possible stable layer at Mbars \citep{Markham24}.  The H/He phase diagram together with a temperature profile determine the He abundance in the 10-100 GPa region. The opacity in the 1-10 kbar region influences the heat flow and the temperature profile underneath. (Normal) modes yield information on the location and thickness of stable layers. Stable layers and albedo influence the heat loss over time and thus the temperatures in the deep interior. The temperature profile throughout and the He abundance profile determine the metallicity-dependent density profile. The gravitational harmonic determine the density profile and thus can rule out many of the assumed metallicity-dependent profiles that made it so far. An observed atmospheric water abundance will rule out others. The surviving profiles place constraints on the global water abundance and thus on the interior structure and formation of Jupiter.

\begin{acknowledgments}
We thank the Juno Science Team and in particular, Ben Idini, Jonathan Lunine, Sushil Atreya, Ravit Helled, and Simon M{\"uller} 
for elucidating discussions.
This work was supported through NASA's Juno Participating Scientist Program under Grant 80NSSC19K1286.
\end{acknowledgments}

\bibliography{jupiterHe24}
\bibliographystyle{aasjournal}

\end{document}